\documentclass[aps,pre,groupedaddress,showpacs,superscriptaddress,twocolumn,amsfonts,amssymb,amsmath,floatfix]{revtex4}

\usepackage{graphicx}
\usepackage{epsfig}

\newcommand{\be}{\begin{equation}}
\newcommand{\ee}{\end{equation}}
\newcommand{\bi}{\begin{itemize}}
\newcommand{\ei}{\end{itemize}}
\newcommand{\ba}{\begin{eqnarray}}
\newcommand{\ea}{\end{eqnarray}}
\newcommand{\bt}{\begin{tabular}}
\newcommand{\et}{\end{tabular}}

\newcommand{\lsim}{\raisebox{-0.5ex}{\ensuremath{\,\,\stackrel{\textstyle <}{\sim}}\,\,}}

\newcommand{\ord}{\mathcal{O}}
\newcommand{\eps}{\varepsilon}
\newcommand{\Del}{\Delta}
\newcommand{\om}{\omega}
\newcommand{\omq}{\omega_{\bf q}}

\newcommand{\Om}{\Omega}
\newcommand{\Omd}{\Omega^\dagger}
\newcommand{\bp}{\mbox{\boldmath$\partial$}}
\newcommand{\bF}{{\bf F}}
\newcommand{\Fex}{F_{\rm ex}}
\newcommand{\bFex}{{\bf F}_{\rm ex}}
\newcommand{\Fzs}{{F^z_s}}

\newcommand{\Omir}{\Om^{\dagger\,\mbox{\tiny irr}}}

\newcommand{\rhosk}{\rho^s_{\bf k}}
\newcommand{\rhosq}{\rho^s_{\bf q}}
\newcommand{\phisk}{\phi^s_{\bf k}}
\newcommand{\phip}{\phi_{\bf p}}
\newcommand{\phisq}{\phi^s_{\bf q}}
\newcommand{\phismk}{\phi^s_{-\bf k}}
\newcommand{\phiq}{\phi_{\bf q}}

\newcommand{\phik}{\phi_{\bf k}}
\newcommand{\rhop}{\rho_{\bf p}}
\newcommand{\hphis}{\widehat{\phi^s}}

\newcommand{\fsr}{f^s({\bf r})}
\newcommand{\fsq}{f^s_{\bf q}}

\newcommand{\fq}{f_{\bf q}}

\newcommand{\bq}{{\bf q}}

\newcommand{\Mir}{M^{\mbox{\tiny irr}}}
\newcommand{\Cir}{C^{\mbox{\tiny irr}}}
\newcommand{\Msqirr}{M^{s, \mbox{\tiny irr}}_{\bf q}}
\newcommand{\Msmqirr}{M^{s, \mbox{\tiny irr}}_{\bf- q}}

\newcommand{\Mqirr}{M^{\mbox{\tiny irr}}_{\bf q}}

\begin{document}


\title{Nonlinear microrheology of dense colloidal suspensions: a mode-coupling theory }


\author{I. Gazuz}
\affiliation{Leibniz-Institut f\"ur Polymerforschung Dresden
  e.~V., Hohe Stra{\ss}e 6, 01069 Dresden, Germany}
\affiliation{Fachbereich Physik, Universit\"at Konstanz, 78457, Konstanz, Germany}

\author{M. Fuchs}
\affiliation{Fachbereich Physik, Universit\"at Konstanz, 78457, Konstanz, Germany}


\date{\today}

\begin{abstract}
A mode-coupling theory for the motion of a strongly forced probe particle in a dense colloidal suspension is presented. 
Starting point is the Smoluchowski equation for $N$ bath and a single probe particle. The probe performs Brownian motion  
under the influence of a strong constant and uniform external force $\Fex$. It is immersed in a dense homogeneous bath of (different) 
particles also performing Brownian motion. Fluid and glass states are considered; solvent flow effects are neglected. 
Based on a formally exact generalized Green-Kubo relation, mode coupling approximations are performed and an integration  
through transients approach applied. A microscopic  theory for the 
nonlinear velocity-force relations of the probe particle in a dense fluid and for the (de-) localized probe in a glass is obtained.
 It extends the mode coupling theory of the glass transition to strongly forced tracer motion and describes active microrheology 
experiments.  A force threshold is identified which needs to be overcome to pull the probe particle free in a glass. For the model  
of hard sphere particles, the microscopic equations for the threshold force and the probability density of the localized probe are 
solved numerically. Neglecting the spatial structure of the theory, a schematic model is derived which contains two types of 
bifurcation,  the glass transition and the  force-induced  delocalization, and
which allows for analytical and numerical solutions. We discuss its phase diagram, forcing effects on the time-dependent
 correlation functions, and the friction increment. The model was successfully applied to simulations and experiments on 
colloidal hard sphere  systems [I.~Gazuz et.~al.,~Phys.~Rev.~Lett.~102, 248302 (2009)], while we provide detailed information 
on its derivation and general properties.
\end{abstract}

\pacs{}

\maketitle


\section[Introduction]{Introduction}\label{intro}

Complex fluids are very common in technological applications as well as in living systems.
Rheology \cite{Larson1999}  can provide deep insight into their mechanical properties, since it 
studies their flow and deformation under external force fields.  While in conventional 
macrorheology \cite{Waigh2005,Squires2010} mechanical experiments in the bulk are performed, in microrheology the diffusive motion of an embedded, 
mesoscopic tracer particle is observed. Microrheology thus has an advantage that also materials can be studied, which are not available
in large amounts. Corresponding experimental techniques were developed during the last years
 \cite{MacKintosh1999,Gisler1998,Gittes1997,Crocker2000}.
They utilize the fluctuation-dissipation 
theorem \cite{Forster1975}, which connects the  linear response of an  observable to external fields with
the corresponding  time-dependent equilibrium correlation function.

To probe the nonlinear properties of the material in a microrheological experiment, the tracer has to be actively pulled by means of an
 external force. Corresponding experiments use magnetic forces 
\cite{Weeks04,Bausch1999} as well as optical tweezers \cite{Furst2003,Meyer2006,Wilson2009} and 
 measure the nonlinear dependence of the probe velocity on the pulling force. 

A typical and ubiquitous nonlinear effect in complex fluids is thinning, i.~e.~the decay of the tracer friction coefficient with 
increasing external force. The theoretical understanding of the thinning effect in microrheology was achieved for  
the case of dilute colloidal suspensions \cite{Brady05} by solving the corresponding two-particle diffusion equation. The results of 
the theory are in good agreement with the 
simulations \cite{Carpen2005} and experiments \cite{Furst2003}.
At larger densities the rheological properties become more complex. If the density exceeds a certain critical value, 
many complex fluids go in to a disordered solid state and exhibit elastic response \cite{Liu1998}.
In this state, yielding is observed, i.~e.~the external field  
must overcome a finite threshold \cite{Hastings03,Reichhardt2010,Fiege2012} in order to produce a flow.
Dense polydisperse colloidal suspensions \cite{Hunter2012} represent one of the simplest model system for such viscoelastic  complex fluids. 
Here, neither an exact solution of the underlying many-particle diffusion equation can be given nor perturbative methods can be applied.
The mode-coupling theory (MCT) proved to be the method of choice for such systems, since it describes the localization of the tracer 
in the cage of its nearest neighbours \cite{Goetze2009} by accounting for the nonlinear backflow effect in a self-consistent manner. 

Recently, a generalization of the standard (quiescent) MCT for the case of nonlinearly pulled tracer was announced 
\cite{Gazuz2008,Gazuz2009}.
The new theory adopts and develops the ideas of the ``integration through transients'' approach to macrorheology 
\cite{Fuchs2002c,Fuchs2009,Brader2012} for the case of microrheology. 
The force-dependent probability density of a localized probe exhibits a bifurcation transition, thus accounting for the yielding effect.
For the tracer friction coefficient (in the fluid state or above the yielding threshold in the jammed state), 
thinning behaviour is observed. 
In \cite{Gazuz2009}, the nonlinear probe velocity-force relations of the schematic model 
 were compared to experiments and simulations. 
Including fluctuations perpendicular to the forcing 
directions, the schematic model was extended in \cite{Gnann2011}, and discussed in detail in \cite{Gnann2012}
The latter model also could be extended \cite{Harrer2012_2} to predict force-induced diffusion \cite{Winter2012} parallel and 
perpendicular to the external force, based on the microscopic memory kernels which we derive here.
The low-force dependence of the tracer probability density was studied in detail \cite{Harrer2012_1}.

While the above-mentioned recent publications focused on comparison of the theory with experiments and simulations 
\cite{Gazuz2009,Gnann2011} and on some of its special aspects \cite{Harrer2012_1} and extensions \cite{Harrer2012_2,Gnann2012}, 
the present paper is intended to provide a comprehensive account of the basics of the  theory. 
We provide the details necessary to understand the derivation of the basic equations and discuss their general properties.
Numerical solutions of the MCT equations and (if available) analytical results are presented and compared with each other for both the 
microscopic version of the theory as well as for the simplified schematic models. For the schematic model, we restrict ourself
to the simplest version (where only fluctuations in the force direction are included) and present its explicit derivation from the microscopic 
theory. Then we discuss the long-time limit, the time dependence 
of the correlators including the asymptotic results and scaling laws at the vicinity of the critical point as well as the resulting friction 
coefficient in detail.
Also, a version of the schematic model (the ``F1-model'') for immobile bath particles is presented, which has not been considered before.
The results here are of technical interest (since the equations are simpler and allow  analytical solutions), 
but might be also of interest in connection with the localization transition in the Lorentz model \cite{Franosch2006}, 
which considers a tagged particle in an array of immobile scatterers.

The paper is organized as follows.
In Sec.~\ref{nonlin_resp}, the generalized  Green-Kubo relation is derived, valid for the nonlinear response to the external force on the tracer.
 From this general relation, we derive the expression for the tracer friction coefficient. The time-dependent transient tracer density correlators 
(being the central quantities in our mode-coupling approach) are then introduced and the mode-coupling equations for them as well as the 
mode-coupling approximation for the tracer friction coefficient are derived.
Sec.~\ref{results} presents results for the hard-sphere system. First, the low-density limit of our theory is studied 
and compared with the exact theory \cite{Brady05}. Then, the bifurcation transition for the long time limit of the tracer density correlator is 
studied in detail.
Sec.~\ref{schem_models}, is devoted to the schematic models.


\section[Theory]{Theory}\label{chap_lin_resp}\label{nonlin_resp}

\subsection{Basic microscopic equations}
\label{smol_eq_sec}

The Smoluchowski equation will provide the basis for all the considerations in this article:
\be\
\partial_t\Psi = \Om\Psi\label{SE},
\ee
where $\Om$ is the  Smoluchowski operator.  
Eq.~(\ref{SE}) describes the time evolution of the $(N+1)$-particle configuration
space probability density $\Psi({\bf r}_1,\,\ldots,{\bf r}_N,{\bf r}_s,t)$ on a
coarse-grained time scale, i.e. it is assumed that the velocity fluctuations
relax much faster than the configurations. 
The particles are colloids performing Brownian motion with diffusion
coefficients $D_i$ in a Newtonian solvent. The particle diffusion coefficients
obey the the Stokes-Einstein relation 
\be\label{st_ein}
D_i = \frac{k_B T}{6\pi\eta\, a_i},
\ee
where $\eta$ is the solvent viscosity and $a_i$ the radius of particle $i$.  
The  colloids are allowed to interact  by means of the potential forces ${\bf F}_i = -\bp_i V({\bf r}_1,\,\ldots,{\bf
  r}_N)$ ($\bp_i$ denotes the partial derivative $\partial/\partial {\mathbf
  r}_i$), whereas the hydrodynamic interactions will be neglected.

\begin{figure}
\includegraphics[scale=0.5]{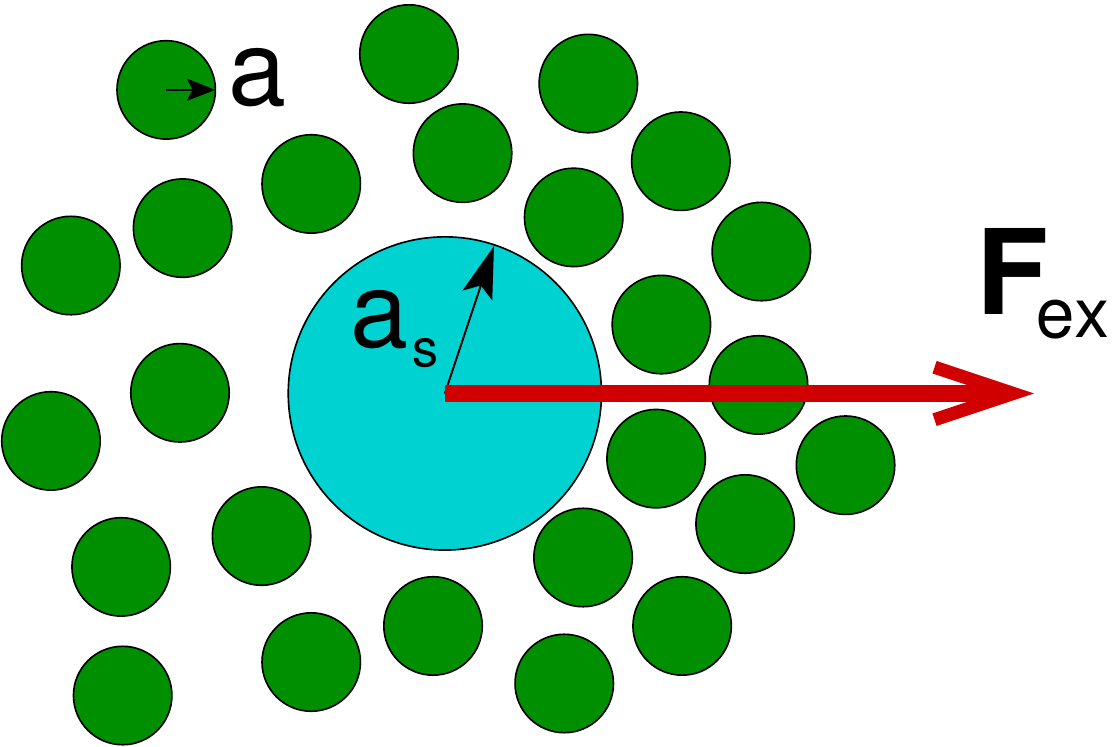}
\caption{\small\label{tracer_bath_fig} (Color online). The colloidal  ``tracer'' particle is pulled through the suspension of ``bath'' particles
by means of the external force $\bF_{ex}$.
The tracer radius is $a_s$, and a bath particle has radius $a$.}
\end{figure}

 We consider a single, distinguished particle (the ``tracer'') with position ${\bf r}_s$ and the  diffusion
 coefficient $D_s$ surrounded by $N$  identical particles (the ``bath''),
 which have the diffusion coefficient $D_0$.
The tracer is pulled by means of the external force  $\bF_{ex}$ (see
Fig.~\ref{tracer_bath_fig}) through the suspension.
The full Smoluchowski operator 
\be\label{smol_op_fex}
\Om = \Om_0 + \Delta\Om
\ee
consists of the unperturbed part
\be\label{om_0}
\Om_0 =  D_0 \sum_{i=1,\,\ldots, N} \bp_i\cdot(\bp_i- \frac{1}{k_BT}\bF_i) + D_s  \bp_s\cdot(\bp_s-\frac{1}{k_BT}\bF_s)
\ee
and the perturbation due to $\bFex$
\be\label{pert_Fex}
\Delta\Om = -\frac{D_s}{k_BT}\bF_{ex}\cdot\bp_s.
\ee
 $\bF_{ex}$ will be assumed to be {\em constant} in space and time. From now on, we set $k_B T =
1$ in the Smoluchowski operator to simplify the notation. For comparisons with experiments or simulations, 
the factor $k_B T$ will be reintroduced.

In the following, equilibrium-weighted averages $ \int d\Gamma \Psi_{eq}\,\ldots$  will appear, which will be
denoted by $\langle \ldots \rangle$, where
\be
\Psi_{eq}=\frac{1}{Z} \, e^{-V(\{{\bf r}_i\},{\bf r}_s)}
\ee
 is the equilibrium distribution of the unperturbed system with the statistical sum $Z$.
We also introduce the usual equilibrium-weighted scalar product, which  is defined as
\be\label{scalar_eq}
\langle A\,|B\rangle = \int d\Gamma \Psi_{eq}\,  A^*(\Gamma)B(\Gamma)
\ee
for two configuration-space observables $A$ and $B$.

\subsection{Nonlinear response to the external force on the tracer}\label{resp_ext_force_sec}\label{ex_sol_sec}

Let us consider the following situation. For times $t<0$, the system is equilibrated and there are no 
external fields. At $t=0$,  the external force on the tracer is switched on, driving the system out of equilibrium. Instead of assuming
that the perturbation is small, like it is done in the linear response theory, we consider the general case of arbitrarily large forces.

With the initial condition
\be
\Psi(t=0) =\Psi_{eq},
\ee
the solution of the Smoluchowski equation can be written down as
\be
\Psi(t) = e^{\Om\,t}\,\Psi_{eq}
\ee
Using the operator identity
\be
e^{\Om\,t} = 1+\int_0^t dt'\, e^{\Om\,t'}\,\Om
\ee
and noting that $\Om\Psi_{eq} = \Delta\Om\Psi_{eq}$, we get
\be\label{delta_Psi}
\Psi(t)  =  \Psi_{eq} + \int_0^t dt'\,e^{\Om\,t'}\Delta\Om\Psi_{eq}
\ee

The mean value of an observable $A({\bf r}_1,\,\ldots,{\bf r}_N)$ at time $t$ is given by:
\be\label{mean_obs}
\langle A (t) \rangle = \int d\Gamma\, \Psi(\Gamma,t) A(\Gamma),
\ee
where $\Gamma$ is a phase space point and the integration goes over the entire phase space. 
Using eq.~(\ref{delta_Psi}),   we obtain
\be
\langle A(t) \rangle =  \langle A \rangle  - D_s\int d\Gamma\,
A(\Gamma)\int_0^t dt'\,e^{\Om\,t'} (\bF_{ex}\cdot\bp_s)\,\Psi_{eq}.
\ee
We note that $\bp_s\,\Psi_{eq} = \bF_s\Psi_{eq}$, introduce 
\be\label{smol_op_fex_back}
\Om^\dagger = \Om_0^\dagger + \Delta\Om^\dagger \,,
\ee
the adjoint of $\Om$ with respect to the unweighted scalar product \cite{Dhont1996}, with 
\be\label{om0_dagger}
\Om_0^\dagger  =  
 \sum_{i=1}^N D_0\, (\bp_i+\bF_i)\cdot\bp_i + D_s\, (\bp_s+\bF_s)\cdot\bp_s
\ee
and $\Delta\Om^\dagger = D_s\, \bF_{ex}\cdot\bp_s$, 
and finally arrive at
\be\label{nonlin_resp_eq}
\langle A \rangle (t) = \langle A \rangle -D_s\,\bF_{ex}\cdot \int_0^t
dt'\,\langle\,\bF_s\,  e^{\Om^\dagger\,t'} A\,\rangle \, .
\ee

Eq.~(\ref{nonlin_resp_eq}) represents the  generalized {\em nonlinear} Green-Kubo relation for
the response of an observable $A$ to the perturbation by the external force on
the tracer. In contrast to the well-known linear response expression, the  full Smoluchowski operator $\Om$ containing the external force, 
 instead of just the unperturbed one  enters eq.~(\ref{nonlin_resp_eq}). 

The presence of an external force renders the operator (\ref{smol_op_fex_back})  {\em nonhermitian} with respect to the 
equilibrium-weighted scalar product (\ref{scalar_eq}). Its adjoint is now given by
\be\label{om_adjoint}
\Om^{adj} = \Om^\dagger_0 - D_s\,\bF_{ex}\cdot(\bF_s + \bp_s)
\ee
(the calculation is presented in \cite{Gazuz2008}).

\subsection[]{Tracer mobility}\label{tr_fric_sec}

The (long-time) tracer  mobility  is defined as
\be\label{mu_def}
\mu_s = \lim_{t\rightarrow\infty}\frac{\langle {\bf v}_s \rangle (t)}{\bF_{ex}},
\ee
where ${\bf v}_s$ is the tracer velocity. In the framework of the Smoluchowski dynamics, where the particle motion is overdamped,
 the tracer velocity ${\bf v}_s = \partial_t {\bf r}_s  =  \Om^\dagger {\bf
   r}_s $ is a function on the configurational space and is given by
\be
{\bf v}_s = \mu_s^0(\bF_{ex} + \bF_s),
\ee
where $\mu_s^0=D_s/(k_B T)$ is the single-particle tracer mobility.
Since $\bF_{ex}$ is given externally and has no dependence on the phase space of the system, the problem reduces to calculating  
the average of the force $\bF_s$ from the bath particles on the tracer. To
determine $\langle F^\alpha_s \rangle$,  the $\alpha$-th component of the vector
$\langle \bF_s \rangle$, we use eq.~(\ref{nonlin_resp_eq}), note that the
equilibrium average $\langle \bF_s \rangle$ vanishes and obtain
\be\label{resp_A}
\langle F^\alpha_s \rangle (t) =   -D_s\bF_{ex}  \int_0^t dt'\,\langle \bF_s\,   e^{\Om^\dagger\,t'}  F^\alpha_s  \rangle \, .
\ee
 Let us introduce the coordinate system such that $\bFex$ points in the positive
$z$-direction. Then we have $\bFex\cdot\bF^s = F_{ex} F^z_s$ in
 eq.~(\ref{resp_A}). Expression (\ref{resp_A}) can be simplified further if we
 employ the rotation symmetry around the $z$-axis, which our system obviously exhibits. After such
 a rotation, the phase space integral in (\ref{resp_A}) should remain the
 same. On the other hand, rotations by angle $\pi$ change the sign of both
the $x$- and the $y$-component of $\bF^s$. This means that the correlators $\langle F^z_s\,   e^{\Om^\dagger\,t'}  F^x_s  \rangle$ and $\langle
F^z_s\,   e^{\Om^\dagger\,t'}  F^y_s  \rangle$ vanish and we are left with
\be\label{Fs_av}
\langle \bF_s \rangle (t) =   - D_s\bF_{ex} 
 \int_0^t dt'\,\langle \Fzs\,   e^{\Om^\dagger\,t'}  \Fzs  \rangle \, .
\ee
As was anticipated in eq.~(\ref{mu_def}), the mean force exerted from the bath on the tracer is
parallel to the external force $\bFex$.

Expression (\ref{Fs_av}) includes the force-force correlator $C(t) = \langle
\Fzs\,   e^{\Om^\dagger\,t'}  \Fzs\rangle$ and leads for the tracer mobility
to the result
\be\label{mob_ct}
\mu_s = \mu_s^0\left(1-D_s\int_0^\infty dt\, C(t)\right).
\ee
For the further use together with the mode-coupling approximations, the
force-force correlator in eq.~(\ref{mob_ct}) should be rewritten in terms of the irreducible
Smoluchowski operator 
\be\label{omir_fsz} 
\Omir =\Omd -
\Fzs\rangle D_s^{-1} \langle \Fzs ,
\ee
 following \cite{CichHess87,Kaw95}. After changing to the Laplace space according to $C(z)=\int_0^\infty dt\, e^{-zt} C(t)$,
\be\label{fcor_lap}
C(z) = \langle \Fzs \frac{1}{z-\Omd} \Fzs \rangle
\ee
 and using the standard operator identity for $A=A_1+A_2$
\be
(z-A)^{-1} = (z-A_1)^{-1} + (z-A)^{-1}A_2(z-A_1)^{-1},
\ee
(with $A=\Omd$, $A_1=\Omir$) for the resolvent in (\ref{fcor_lap}), one  obtains the expression
\be\label{c_cir}
C(z) = \frac{\Cir(z)}{1+D_s\Cir(z)}
\ee
for the force-force correlator in terms of the irreducible one
$\Cir(z)=\langle\Fzs\,(z-\Omir)^{-1} \Fzs\rangle$. Exploiting the relation
(\ref{c_cir}) for $z=0$ leads us to the desired result
\be\label{mu}
\mu_s  =  \frac{\mu_s^0}{1 + D_s\int_0^\infty dt\, \Cir(t)}
\ee
for the tracer mobility in terms of the irreducible tracer force
autocorrelation function $\Cir(t)$, which in the time domain is given by
\be\label{Cf_irr}
\Cir(t) = \langle \Fzs\,   e^{\Omir\,t}  \Fzs  \rangle
\ee

Eq.~(\ref{mu}) allows a simple interpretation if one introduces the  tracer
{\em friction coefficient} $\zeta_s = 1/\mu_s$. The friction coefficient is  given
by the sum
\ba\label{zeta}
\zeta_s & = & \zeta_s^0 + \Delta\zeta_s,\\
\label{del_zeta}
\Delta\zeta_s & = & \int_0^\infty dt\,\Cir(t),
\ea
of the ``bare'' (single-particle) tracer friction coefficient due to the solvent, given by $\zeta_s^0 = 1/\mu_s^0$
and the {\em  increment} $\Delta\zeta_s$  due to interactions with the bath
particles. 
Because we are interested in dense dispersions where the friction is highly increased beyond the solvent one, Eq.~(\ref{zeta}) provides a more 
secure route to approximations than Eq.~(\ref{mob_ct}). In Eq.~(\ref{zeta}) slow force fluctuations contribute to an increased friction.
 In Eq.~(\ref{mob_ct}),  instantaneous and retarded velocity fluctuations need to cancel in order to yield a reduced mobility. 
The mode coupling approximations to be performed 
 later set up a self-consistent set of equations for slow fluctuations which is better tailored to  Eq.~(\ref{zeta}) than to Eq.~(\ref{mob_ct}).
Recent simulations of a forced probe in a bath of noninteracting bath particles support to approximate $\Cir$ instead of $C$ \cite{Zia2010}. 
Even though the bath particles do not interact among themselves, the collisions with the probe particle
induce correlations in the velocity but not (or to lesser extent) in the force fluctuations. The friction increment increases linearly with bath
density, as expected from independent collisions of the non-interacting bath particles with the probe. This expected behavior, however, does not
hold for the mobility change, which varies more rapidly with bath density \cite{Zia2010}.

After these formally exact manipulations, approximations are now required in order to evaluate the irreducible force correlation function. 
Low density approximations have been performed \cite{Brady05}, and we will use mode 
coupling approximations to address high packing fractions close  to the colloidal glass transition.

\subsection{Transient tracer density fluctuations}\label{phiskFex}

The important quantities in the mode-coupling approach are the tracer and the bath densities
\ba
 \rho^s({\bf r}) & = &\delta({\bf r}-{\bf r}_s),\\
 \rho({\bf r}) & = &\sum_{i=1}^N\delta({\bf r}-{\bf r}_i).
\ea

With the convention for the Fourier transform of a function $X({\bf r})$ to be
\be\label{FT_def}
X({\bf q})  =  \int d{\bf r}\, e^{i\, {\bf q}\cdot {\bf r}}\, X({\bf r}),
\ee
implying
\be\label{FTback_eq}
X({\bf r})  = \frac{1}{(2\pi)^3} \int d{\bf q}\, e^{-i\, {\bf q}\cdot {\bf r}}\, X({\bf q})
\ee
for the back-transform, we have
\ba
\rhosq & = & e^{i\,{\bf q}\cdot{\bf r}_s} \\
\rho_{\bf q} & = & \sum_{i=1}^N e^{i\,{\bf q}\cdot{\bf r}_i}.
\ea
for the tracer and the bath density modes.

\subsubsection{General properties}\label{phisq_gen}

The first question to clarify concerns the time-dependent correlator 
\be\label{rhosk_rhosk'}
\langle \rhosq e^{\Omd\,t}  \rho^s_{{\bf q}'}  \rangle
\ee  
of  two  tracer density modes. For which pairs of wavevectors ${\bf q}$, ${\bf q}'$ is it  nonzero ? 

For the  case of an isolated system, translational invariance implies that after all particle positions 
have been shifted
\be\label{coord_shift}
\Gamma\rightarrow\Gamma'
\ee 
with ${\bf r}_s  \rightarrow  {\bf r}_s + {\bf a}$,
${\bf r}_i  \rightarrow  {\bf r}_i + {\bf a}$
($i=1,\ldots,\,N$), the average (\ref{rhosk_rhosk'}) should remain the same. Since the shift introduces the prefactor 
$e^{i\,({\bf q} + {\bf q}')\cdot {\bf a}}$ in eq.~(\ref{rhosk_rhosk'}) and the vector ${\bf a}$ can be arbitrary, the condition 
\be\label{k_eq_mk'}
{\bf q} = - {\bf q}'
\ee
follows.

Our driven system is translationally invariant as well, since we
assume the external force $\bF_{ex}$ to be space and time independent and the Smoluchowski operator
(\ref{smol_op_fex_back}) does not change after the shift (\ref{coord_shift}). Thus, the argumentation used for the isolated systems and
thus the condition (\ref{k_eq_mk'}) remains.

We can thus introduce the usual notation
\ba\label{phisq_def}
\phi^s_{\bf q}(t) & = & \langle  \rho^{s*}_{\bf q} e^{\Omd\,t} \rho^s_{{\bf q}}\rangle,\\
\phi_{\bf q}(t) & = & \frac{1}{N S_q} \langle  \rho^*_{\bf q} e^{\Omd\,t}
\rho_{{\bf q}} \rangle\, 
\ea 
for the tracer and the bath density mode correlators, where $S_q$ is the bath static structure factor
\be\label{Sk_def}
S_q = \frac{1}{N}\, \langle  \rho^*_{{\bf q}} \, \rho_{\bf q}  \rangle.
\ee

The Fourier back transform of $\phisq(t)$:
\be\label{phisr_def}
\phi^s({\bf r}) (t) = FT^{-1}[\phisq(t)].
\ee
 is the probability density of finding the tracer at the point ${\bf r}$  in space at time  $t$ with the initial condition that
at time $t=0$ it was localized at the origin. The derivation is given in  Ref.~\cite{Hess1983}, where an isolated system is considered.
Since no special properties of the Smoluchowski operator for isolated systems are used in Ref.~\cite{Hess1983},
the derivation is valid  also  for our case.

The general condition for the Fourier back transform (\ref{phisr_def}) to be real is
\be\label{phimink}
\phi^s_{-{\bf q}}(t) = \phi^{s\,*}_{\bf q}(t).
\ee
We can easily see that  the property (\ref{phimink}) is indeed fulfilled by
$\phisq$ if we use it's definition (\ref{phisq_def})  and the fact that
the operator $e^{\Omd\, t}$ is linear and real (since it contains derivatives with respect to phase space coordinates multiplied 
by real numbers). So,
\ba\nonumber
\phi^{s\,*}_{\bf q}(t) = \langle  \left(e^{i\,{\bf q}\cdot{\bf r}_s}\, e^{\Omd\,t}\, e^{-i\,{\bf q}\cdot{\bf r}_s}\right)^*\rangle = \quad\quad\quad\quad\\
= \langle e^{-i\,{\bf q}\cdot{\bf r}_s}\, e^{\Omd\,t}\, e^{i\,{\bf q}\cdot{\bf r}_s}  \rangle = \phi^s_{-\bf q}(t)
\ea

\subsubsection{Zwanzig-Mori equations}

The Zwanzig-Mori projector operator formalism allows one to express the fluctuations of a given observable in
terms of the corresponding memory kernel. After introducing the projectors
$P_A = A\rangle\langle A^*$, $Q_A = 1-P_A$ for an observable $A$ (we assume
that $\langle A^* A\rangle = 1$ for simplicity), the Laplace
transform $C(z)=\int_0^\infty dt\, e^{-zt} C_A (t)$ of its equilibrium time
correlation function $C_A (t) = \langle  A^* e^{\Omd\,t} A\rangle $ can be
written as \cite{Forster1975} 
\be\label{mori_A}
C_A (z) = \frac{1}{z + \om_A - M_A(z)},
\ee
where the frequency $E_A$ and the memory function $M_A(z)$ are given by
\ba
\om_A & = & - \langle  A^* \Omd A\rangle, \\
M_A(z) & = & \langle  A^* \Omd\, Q_A \frac{1}{z - Q_A \Omd Q_A} Q_A \Omd  A\rangle.
\ea
In the time domain, eq.~(\ref{mori_A}) corresponds to
\be
\partial_t\,C_A(t) = -\om_A\,C_A(t) + \int_0^t dt' M_A(t-t')\,C_A(t').
\ee
For dissipative systems, like for the case of our system described by the
Smoluchowski operator, a second projection step is needed \cite{CichHess87,Kaw95}. To this end, one introduces the
irreducible Smoluchowski operator  
\be\label{omir_A}
\Omir = Q_A\left(\Omd - \Omd A\rangle\langle A^*\Omd A  \rangle^{-1}\langle A^* \Omd\,\right) Q_A.
\ee
and gets the representation
\be
M_A(z) = \frac{\Mir_A(z)}{1+\om_A^{-1}\Mir_A(z)}
\ee
for the memory function in terms of the irreducible memory function
$\Mir_A(z)$ given by
\be
\Mir_A(z)  =  \langle  A^* \Omd\, Q_A \frac{1}{z - \Omir} Q_A
\Omd  A\rangle,
\ee
 which time evolution is generated by  $\Omir$: $\Mir_A(t)  =  \langle  A^*
 \Omd\, Q_A e^{\Omir t} Q_A \Omd  A\rangle$.
For the irreducible memory equation, one obtains in the time domain
\be
\partial_t\,C_A(t) = -\om_A\,C_A(t) - \frac{1}{\om_A}\int_0^t dt' \Mir_A(t-t')\,\partial_t'\,C_A(t').
\ee

The procedure of expressing the equilibrium correlation
functions in terms of  memory kernels  sketched above, was originally proposed for an isolated system evolving
with the unperturbed operator $\Omd_0$. We apply it to the correlators evolving
with the nonhermitian operator $\Omd$  even though the mathematical
conditions and justifications are unknown at present. This procedure is based
on the conjecture that the algebraic structure of the Zwanzig-Mori equations
together with the mode coupling approximations, necessary in the latter steps
to evaluate them, capture the mathematical bifurcation describing the
delocalization of the probe under strong force. At present this conjecture can
only be tested by formulating the theory and considering its results in comparisons with data and formal symmetry requirements.

After these considerations, we are in a position to write down the memory
equation for $\phisq(t) = \langle \rho^{s*}_{\bf q} e^{\Omd\,t} \rho^s_{{\bf q}}\rangle$:
\be\label{memeq}
\partial_t\,\phisq(t) = -\om_{\bf q}\,\phisq(t) - \frac{1}{\omq}\int_0^t dt' \Msqirr(t-t')\,\partial_{t'}{\phi}^s_{\bf q}(t'),
\ee
with
\ba
\om_{\bf q} & = & - \langle \rho^{s*}_{\bf q}\,\Omd\,\rho^s_{\bf q}  \rangle,\\\label{msqirr}
\Msqirr(t) & = &  \langle \rho^{s*}_{\bf q}\,\Omd\, Q^s\,e^{\Omir t}\,
Q^s\,\Omd\,\rho^{s}_{\bf q}  \rangle \\\label{omir_rhosq}
\Omir & = & Q^s\left( \Omd -\Omd\,\rhosq\rangle \,\omq^{-1}\langle \rho^{s*}_{\bf q}\, \Omd\, \right) Q^s
\ea
and the projector
\be\label{qsps}
Q^s  =  1 -  \rho^s_{\bf q}\rangle\langle \rho^{s*}_{\bf q} \,.
\ee
Applying $\Omd$ to $\rhosq$ yields
\ba\label{omrhoq}
\Omd \,\rhosq = D_s\,(\bp_s^2 + (\bF_s +\bF_{ex})\cdot\bp_s) \, e^{i\,{\bf q}\cdot {\bf r}_s}    \quad\quad\\\nonumber
= D_s\, (-q^2 + i\,{\bf q}\cdot(\bF_s + \bF_{ex}))\,\rhosq\,,
\ea
so for the frequency $\omq$ we obtain
\be\label{omq_Fex}
\omq = D_s\, ( q^2 -i\,{\bf q}\cdot\bF_{ex}).
\ee
The fact that  $\omq=- \langle \rho^{s}_{\bf q}\,|\,\Omd\,|\,\rho^s_{\bf q}  \rangle$ turns out to be complex  is obviously the consequence of
the mentioned nonhermiticity of the operator $\Omd=\Omd_0 + D_s\,\bF_{ex}\cdot\bp_s$ 
with respect to the equilibrium-weighted scalar product (\ref{scalar_eq}).

We would like to discuss now the relationship between the irreducible operator
introduced in eq.~(\ref{omir_fsz}) for the force-force correlator (call it $\Omir(\Fzs)$) and the one
introduced in eq.~(\ref{omir_rhosq})  for the tracer density modes (call it $\Omir(\rhosq)$). To this end, we set $\bFex = 0$,
 go to the limit ${\bf q} \rightarrow 0$ in eq.~(\ref{omir_rhosq}) and employ relations (\ref{omrhoq}), (\ref{omq_Fex}).  Then, we readily see
 that 
\be
\lim_{{\bf q}\rightarrow 0} \Omir (\rhosq) = \Omir (\Fzs) \quad
(\mbox{at}\ \  \Fex = 0)
\ee
 holds. It is not surprising,
since the $({\bf q}\rightarrow 0, z\rightarrow 0)$ limit for the tracer density fluctuations is
related to the tracer diffusion and in the absence of the external force,  the
well-known relation  \cite{Naegele1998b,Fuchs1998}
\be
D^{L}_s = \frac{D_s}{1+\lim_{{\bf q}\rightarrow 0,z\rightarrow 0}\left(
  \Msqirr({\bf  q},z)/\omq \right)} 
\ee
for the long-time tracer diffusion coefficient $D^{L}_s$  implies the {\em Einstein relation} 
\be
D^{L}_s=\mu_s \quad  (\mbox{at}\ \  \Fex = 0)
\ee
 connecting $D^{L}_s$  with the long time tracer mobility $\mu_s$ considered in
 the last section. We see that the irreducible memory function plays the role of a generalized friction kernel.

\subsection{Mode-coupling approximations}
\subsubsection{The memory function}

In order to obtain a self-consistent equation for $\phisq$ from the memory
equation (\ref{memeq}), the irreducible memory function (\ref{msqirr}) is treated using the standard
approximation steps of the mode-coupling theory \cite{Goetze2009}. To this end,
first, the projectors onto the space spanned by the tracer-bath density products
\be\label{p2s}
P^s_2 = 
\sum_{{\bf k}, {\bf p}, {\bf k}', {\bf p}' } \rho^s_{\bf k} \rho_{\bf p}\rangle\, g({\bf k}, {\bf p}, {\bf k}', {\bf p}') \,
 \langle\rho^{s*}_{\bf k'}\rho^*_{\bf p'}
\ee
 are introduced, where the normalization matrix $g$  has to obey the condition
\be\label{cond_g}
\sum_{{\bf k}', {\bf p}'} \langle \rho^s_{\bf k'} \rho_{\bf p'}\, |\, \rho^{s}_{\bf k} \rho_{\bf p} \rangle\,  g({\bf n}, {\bf m}, {\bf k}', {\bf p}') 
 = \delta_{{\bf n},{\bf k}}\,\delta_{{\bf m},{\bf p}}
\ee
which requires an arbitrary vector $|\, \rho^s_{\bf k} \rho_{\bf p} \rangle$ from the space of the tracer and bath density products to be
left invariant upon application of $P^s_2$.

As the first approximation, the ``fluctuating forces''
$Q^s\,\Omd\,\rho^{s}_{\bf q}$ in eq.~(\ref{msqirr}) are replaced by the projected ones:
\be\label{mcapp}
\Msqirr  \approx  \langle \rho^{s*}_{\bf q}\,\Omd\, Q^s\,P^s_2\, e^{\Omir t }\,P^s_2\, Q^s\,\Omd\,\rho^{s}_{\bf q} \rangle.
\ee
In the next approximation step, the four-point 
correlators $\langle\rho^{s*}_{\bf k}\rho^*_{\bf p}\, e^{\Omir t } \,
\rho^s_{{\bf k}'} \rho_{{\bf p}'}\rangle$ appearing in eq.~(\ref{mcapp}) are  factorized into products of the two-point ones \cite{Goetze91}:
\ba\label{fact_appr}
\langle \rho^{s*}_{\bf k}\rho^*_{\bf p}\, e^{\Omir t }\, \rho^s_{{\bf k}'}
\rho_{{\bf p}'}\rangle \quad\quad\quad\quad\quad\quad\quad\quad\quad\quad\quad\quad\quad \\\nonumber \approx 
\delta_{{\bf k},{\bf k}'} \, \delta_{{\bf p},{\bf p}'}\, \langle  \rho^{s*}_{\bf k} e^{\Omd\,t} \rho^s_{{\bf k}'}  \rangle\,
\langle  \rho^*_{\bf p} e^{\Omd\,t} \rho_{{\bf p}'} \rangle \\ \nonumber
 =  \delta_{{\bf k},{\bf k}'} \, \delta_{{\bf p},{\bf p}'}\,
 \phisk(t)\,N S_p\, \phip(t)    .
\ea
Note also that as part of the approximation, the irreducible operator on the left hand side of eq.~(\ref{fact_appr}) was 
replaced by the normal one on the right hand side.

It is easy to see that for $t=0$, the factorization approximation
(\ref{fact_appr})  becomes exact and allows us to calculate the normalization matrix $g$ from the condition (\ref{cond_g}),
which then reads
\be
\sum_{{\bf k}', {\bf p}'}  \delta_{{\bf k}',{\bf k}}\,\delta_{{\bf p}',{\bf p}} N S_p\, g({\bf n}, {\bf m}, {\bf k}', {\bf p}')
  = \delta_{{\bf n},{\bf k}}\,\delta_{{\bf m},{\bf p}},
\ee
leading to the result
\be\label{norm_matrix}
g({\bf n}, {\bf m}, {\bf k}, {\bf p}) = \frac{1}{N S_p} \delta_{{\bf n},{\bf k}}\,\delta_{{\bf m},{\bf p}}.
\ee


The only missing parts appearing after the application of the projectors
$P^2_s$ in the expression (\ref{mcapp}) are now the static averages of the
form  
\be\label{rhokp_q_om_rhosq}
\langle \rho^{s*}_{{\bf k}} \rho^*_{{\bf p}}\, Q^s \Omd\rhosq\rangle
\ee
 and
\be\label{rhosq__om_q_rhokp}
\langle \rho^{s*}_{\bf q}\, \Omd\, Q^s  \rho^s_{{\bf k}}\rho_{{\bf p}} \rangle.
\ee

 First, we notice that since $\Omd$ is not self-adjoint, the averages
 (\ref{rhokp_q_om_rhosq}), (\ref{rhosq__om_q_rhokp}) are not complex
 conjugated of each other, as it would be the case for an isolated system.
Using  eqs.~(\ref{qsps}), (\ref{omrhoq}),  for the term (\ref{rhokp_q_om_rhosq}) we obtain
\ba\nonumber
\frac{1}{D_s}\langle \rho^{s*}_{\bf k} \rho^*_{\bf p}\, Q^s\Omd\rhosq \rangle
 = 
\langle \rho^{s}_{\bf q - k}\, \rho^*_{\bf p}\, (-q^2 + i\,{\bf q}\cdot(\bF_s +
\bF_{ex}))\rhosq\rangle \\ \label{first_av_appb}
 - \,
 \langle \rho^{s}_{\bf q - k}\, \rho^*_{\bf p}  \rangle \,
  (-q^2 + i\,{\bf q}\cdot(\bF_s + \bF_{ex}))\,. \quad\quad\quad\quad
\ea
The nontrivial terms in the above expression are either of the form of the tracer-bath static structure factor 
\be\label{sps}
S^s_p = \,\langle \rho^{s*}_{\bf p} \, \rho_{{\bf p}}\rangle
\ee 
or the term of the form   $\langle \bF_s\, \rhosk \rhop \rangle$. The latter can be
reduced to the tracer-bath static structure factor 
by means of partial integration
\ba\label{red_av_Fs}\nonumber
\langle \bF_s\, \rhosk \rhop \rangle = \frac{1}{Z} \int d\Gamma\, (\bp_s e^{-V}) \rhosk \rhop = \quad\quad\quad\quad\quad\quad\quad \\
   =  - \frac{1}{Z}\int d\Gamma\,e^{-V} \bp_s (\rhosk \rhop)  
  =  - i {\bf k}\, \delta_{{\bf k}+{\bf p},0}\, S^s_p\,. \quad\quad
\ea
Using this relation, one readily obtains the result
\be\label{rhoskp_qs_om_res}
\langle \rho^{s*}_{\bf k} \rho^*_{\bf p}\, Q^s\Omd\,\rhosq\rangle = 
  \delta_{{\bf q},\, {\bf k} + {\bf p}}\, D_s S^s_p\,({\bf q}\cdot{\bf p}) 
\ee

To calculate the remaining average (\ref{rhosq__om_q_rhokp}), the fastest way
is to we use the adjoint of $\Omd$:
\be
\langle \rho^{s*}_{\bf q}\, \Omd Q^s\, \rho^s_{\bf k}\rho_{\bf p} \rangle =
 \langle(\Om^{adj} \rho^{s*}_{\bf q}) Q^s\, \rho^s_{\bf k}\rho_{\bf p} \rangle
\ee
and its action on the conjugated tracer density mode (see eq.~(\ref{om_adjoint})):
\be\label{eq_B22}
\Om^{adj} \, \rho^{s*}_{\bf q} =
D_s\, (-q^2 - i\,{\bf q}\cdot(\bF_s - \bF_{ex}) - \bF_{ex}\cdot\bF_{s} )\,\rho^{s*}_{\bf q}.
\ee
The calculations go in the same way as for the term (\ref{rhokp_q_om_rhosq})
and we omit the details except for the fact that compared to the expression (\ref{omrhoq}), an additional  term $\bF_{ex}\cdot\bF_{s}$
is present in the brackets in eq.~(\ref{eq_B22}), so that the final result also contains $\bFex$:
\be\label{rhoq_om_qs_res}
\langle \rho^{s*}_{\bf q}\, \Omd Q^s\, \rho^s_{\bf k}\rho_{\bf p} \rangle 
   =   \delta_{{\bf q},\, {\bf k} + {\bf p}}\,D_s S^s_p\, ({\bf q}\cdot{\bf p} - i\, \bF_{ex}\cdot{\bf p} )\,.
\ee

After collecting the terms together, using relations (\ref{mcapp}),
(\ref{fact_appr}), (\ref{norm_matrix}), (\ref{rhoskp_qs_om_res}) and
(\ref{rhoq_om_qs_res}), the final mode-coupling expression for the memory
function reads
\be\label{memfq_fex}
\Msqirr (t) =  \sum_{{\bf k}+ {\bf p} =\, {\bf q}} \frac{D^2_s {S_p^s}^2}{N S_{p}}\, 
{\bf q}\cdot{\bf p}\left({\bf q}\cdot{\bf p} - i\,\bF_{ex}\cdot {\bf p} \right) \,
\phi^s_{{\bf k}}(t)\phi_{{\bf p}}(t).
\ee

\subsubsection{The bulk dynamics}\label{bulk_dyn}

So far, nothing was said about the fluctuations of the bulk density modes
$\phi_{{\bf q}}(t)$. The simplest reasonable assumption for these is, to
neglect the effect of the external force. In the thermodynamic limit, which is
considered in this article, this assumption is justified, since the effect of
the tracer on the bath will be to perturb its neighbourhood only locally. This
effect will be included in our theory and can be described by the tracer and
bath density mode products $\rhosq \rho_{\bf q}$, which
are the Fourier space counterparts of the relative bath-tracer density
$\rho({\bf r} - {\bf r}_s)$.

So, we assume the bulk bath dynamics to be unaffected by the external force. The memory
equation for the bulk dynamics reads
\be\label{memeq_phiq}
\tau_q\,\partial_t\phiq(t) + \phiq(t) + \int_0^t dt'\, \Mir_{\bf q}(t-t')\,\partial_{t'} \phiq(t') =0 \,
\ee
with 
\be\label{tauq}
\tau_q = S_q / (D_0 q^2)\,.
\ee

The standard equilibrium MCT expression for the irreducible bulk memory function is 
\be
\Mir_{\bf q}(t) = \frac{1}{2 q^4}\sum_{{\bf k}+ {\bf p} 
=\, {\bf q}} n\, S_q S_k S_p\,\left({\bf q}\cdot ({\bf k}c_k + {\bf  p}c_p)\right)^2 \phik(t)\phip(t),
\ee
where $n=N/V$ denotes the number density of the bath particles and $c_q$ is the Ornstein-Zernike direct correlation function: 
\be
S_q=1/(1-nc_q).
\ee
Results from these well studied MCT equations \cite{Goetze91,Goetze2009} will be used in the following whenever properties of the unperturbed 
bath particles are required. As most important result let us recall already here that MCT predicts a glass transition of colloidal dispersions at high 
concentrations when density fluctuations do not relax completely.

\subsubsection{The force-force correlator}
\label{mct_fscor_sec}
In order to obtain the mode-coupling approximation for the irreducible force-force
correlation function (\ref{Cf_irr}), which enters the expression (\ref{zeta}) for the
tracer friction coefficient, we use the usual MCT strategy and substitute the
 force ${\bf F}_s$ by the one projected to the tracer-bath pair density modes:
\be\label{mcf}
\Cir(t) = \langle \Fzs\, e^{\Omir t }\,\Fzs \rangle \approx \langle \Fzs\,P^s_2\, e^{\Omir t }\,P^s_2\,\Fzs \rangle,
\ee
where $P^s_2$ is given by eq.~(\ref{p2s}). 

Eq.~(\ref{mcf}) is the simplest possible approximation of the mode-coupling type for the
tracer force autocorrelator, since
a product of at least one bath and one tracer density mode is needed. This is
due to the fact that the equilibrium average with the 
tracer force  $\langle \Fzs\,\ldots  \rangle$ appearing in eq.~(\ref{mcf}) is zero both for a single tracer and for a 
single bath density mode.

The calculation is completely analogous to that of the memory function in the
previous paragraph. One uses the factorization approximation (\ref{fact_appr})
as well as the relations (\ref{norm_matrix}), (\ref{sps}) and
(\ref{red_av_Fs}) to obtain
\be\label{corfs}
\Cir(t) \approx  \sum_{\bf k} \frac{1}{N S_k}k_z^2\, {S_k^s}^2\,\phi^s_{\bf k}(t)\phik(t).
\ee

\subsection[Reality of the observable averages]{Reality of the observable averages}\label{real_obs_sec}

In order to check, whether our MCT approximations preserve the reality of
observable quantities, we choose the  friction coefficient as the typical
example. The MCT expression (\ref{corfs}) for the irreducible force-force
correlator $\Cir(t)$  enters the  eq.~(\ref{del_zeta})  for the friction
coefficient increment under the time integral.

Eq.~(\ref{corfs}) contains the sum (over all  ${\bf k}\in {\mathbb R}^3$) over  $\phisk$ multiplied with real and rotationally 
invariant (in the ${\bf k}$-space) factors. The similar structure arises also if one makes
mode-coupling approximations for other (tracer-related) correlators, since the dynamic
part is given by the factorized four-point tracer-bath correlators and the
response quantity-specific  part comes in via the different static k-dependent ``vertices''.

In order for the friction coefficient to be real,  it  suffices to show that the MCT approximation for the tracer density
correlator $\phisk$  fulfills the condition (\ref{phimink})  since then the imaginary parts in the ${\bf k}$-sum cancel.

In the mode-coupling equation (\ref{memeq}) for $\phisq$,
 the memory function (\ref{memfq_fex}) couples the correlator for the wave vector ${\bf q}$ to the correlators 
for all the other wave vectors, one has to consider eq.~(\ref{memeq}) as a
system of coupled equations for the set of all ${\bf q}$.

From a purely mathematical standpoint, it can have different solutions depending on the initial values $\phisq(t=0)$ which in general 
might not fulfill the condition (\ref{phimink}). 
But in our physical problem 
\be\label{init_cond_phisk}
\phisq(t=0)=1
\ee
holds so that (\ref{phimink}) is fulfilled for the initial values.

We show now that the assumption that (\ref{phimink}) is valid for $t>0$ does not contradict the system of equations (\ref{memeq}). For this purpose 
we look at the equation for $-{\bf q}$:
\be\label{eqmink}
\partial_t\,\phi^s_{-{\bf q}}(t) = -\om_{-\bf q}\,\phi^s_{-{\bf q}}(t) -
\int_0^t dt' \Msmqirr(t-t')\,\partial_{t'}{\phi}^s_{-\bf q}(t'),
\ee
where
\be\label{ommink}
\om_{-\bf q} = D_s (q^2 + i\,{\bf q}\cdot\bF_{ex}) = \om_{\bf q}^*
\ee
and 
\be\label{memmink}
\Msmqirr(t) = \sum_{{\bf k}'+ {\bf p}' =\, {-\bf q}} \frac{{S_{p'}^s}^2}{N S_{{\bf p}'}}\, 
{\bf q}\cdot{\bf p}'\,({\bf q}\cdot{\bf p}' + i\,\bF_{ex}\cdot {\bf p}')\,
\phi^s_{{\bf k}'}(t)\,\phi_{{\bf p}'}(t)
\ee
Assumption (\ref{phimink}) yields:
\be\label{Mmink}
\Msmqirr(t) = (\Msqirr)^*(t).
\ee
To prove this, we notice that in the expression (\ref{memfq_fex}) for the memory function
every term for a certain pair of wave vectors $({\bf k},\, {\bf p})$ is complex conjugated 
with the term in expression (\ref{memmink}), 
 corresponding to the pair of wave vectors $({\bf k}',\,{\bf p}')$ with ${\bf k}' = -{\bf k}$, ${\bf p}' = -{\bf p}$:
\ba\nonumber
  \frac{{S_p^s}^2}{N S_{p}}\,
{\bf q}\cdot(-{\bf p})\,\left({\bf q}\cdot(-{\bf p}) + i\,\bF_{ex}\cdot (-{\bf p}) \right)
\phi^s_{-{\bf k}}(t)\,\phi_{-{\bf p}}(t)\quad\quad \\\nonumber
 =  \frac{1}{N S_{p}}\, {S_p^s}^2\,\,\,
{\bf q}\cdot{\bf p}\,\left({\bf q}\cdot{\bf p} + i\,\bF_{ex}\cdot {\bf p} \right)
\phi^{s\,*}_{{\bf k}}(t)\,\phi_{{\bf p}}(t)\quad\quad\quad\quad \\
 =  \left(\frac{1}{N S_{p}}\, {S_p^s}^2\,\,
{\bf q}\cdot{\bf p}\,\left({\bf q}\cdot{\bf p} - i\,\bF_{ex}\cdot{\bf p} \right)
\phi^{s}_{{\bf k}}(t)\,\phi_{{\bf p}}(t)\right)^*\quad\quad
\ea
So, given that (\ref{phimink}) holds, (\ref{Mmink}) holds also. On the other side, if we use (\ref{phimink}) 
in eq.~(\ref{eqmink}), we get:
\be
\partial_t\,\phi^{s\,*}_{\bf q}(t) = -\om_{-\bf q}\,\phi^{s\,*}_{\bf q}(t) - \int_0^t dt' \Msmqirr(t-t')\,\partial_{t'}\phi^{s\,*}_{\bf q}(t'),
\ee
and this equation is the complex conjugated of eq.~(\ref{memeq}) for $\phisq$ due to relations (\ref{ommink}) and (\ref{Mmink}).

These considerations show that the condition (\ref{phimink}) is consistent
with the mode-coupling equations (\ref{memeq}), so that  we can state that the
condition for the reality of the Fourier back 
transform  at least can be {\em imposed} on the
set of mode-coupling equations  for the tracer density mode correlators. 
The latter operation is actually analogous to assuming the correlators to be isotropic for the case of quiescent suspensions
 as it has been always (implicitely) done before. If an external force is present, the corresponding condition (\ref{phimink}) following
from the symmetry is less intuitive and was thus discussed here in somewhat detail.

 A rigorous proof that (\ref{phimink}) will hold for $t>0$, provided that the initial conditions (\ref{init_cond_phisk})
hold, however, was not given here. Such a proof  would require 
the explicit construction of the solution. The numerical results for hard
sphere glasses in Sect.~\ref{long_time_limit} violate (\ref{phimink})  for strong forces in a narrow angle
of directions around the one perpendicular to the external force. This will be 
discussed in more detail in the later section and could indicate a break-down
of the theory for very large forces. Yet, we continue on the assumption that our qualitative results are not affected.


\section{Results for the hard sphere system}\label{results}

Here we apply the microscopic formalism developed in the last Section, to the colloidal
hard sphere system. The tracer radius $a_s$ can be different from that of the
bath particles $a$ (see Figure \ref{tracer_bath_fig}). The control  parameters
are then  the ratio $\alpha=a_s/a$ of the radii of the tracer  and the bath particles
and the volume fraction of the bath particles
$\varphi = \frac{4}{3}\,\pi a^3\,n$.


\subsection[]{Low density limit}\label{ldlimit_nonlin}

First, we would like to calculate the tracer friction coefficient increment
$\Delta \zeta_s$ in the limit $\phi\rightarrow 0$ of vanishing volume fraction
of the bath particles.  
We use the relation (\ref{del_zeta}) for $\Delta \zeta_s$ in terms of the force correlator $\Cir(t)$  together with the MCT
approximation (\ref{corfs}) for $\Cir(t)$ and obtain 
\be\label{delzeta_mct}
\Delta\zeta_s  =  \frac{1}{(2\pi)^3\,n} \int d{\bf k}\,  k_z^2\, \frac{{S_k^s}^2}{S_k}\,\int_0^\infty dt\,
     \phisk(t)\phik(t)\, .
\ee
Note that the  ${\bf k}$-sum in (\ref{corfs})
was changed to the integral over the ${\bf k}$-space: $\sum_{\bf k} \rightarrow V/(2\pi)^3 \int d{\bf k}$.

The static structure factors $S_k^s$, $S_k$ entering eq.~(\ref{delzeta_mct}) can be
easily calculated to the leading order in $\phi$.  We start with the tracer-bath structure factor:
\be\label{Sks_low_appr}
S_k^s = \langle\, \sum_{i=1}^N e^{i\,{\bf k}\cdot({\bf r}_i-{\bf r}_s)}  \rangle \approx
N\,\langle  e^{i\,{\bf k}\cdot({\bf r}_1-{\bf r}_s)} \rangle.
\ee
The approximation made here consists in considering the system as a conglomerate of independent two-particle
clusters. The two-particle structure factor $\langle
e^{i\,{\bf k}\cdot({\bf r}_1-{\bf r}_s)} \rangle$ can be reduced to an
elementary integral over ${\mathbb R}^3$  and  yields a Bessel function:
\be\label{Sks_low_final}
S_k^s  \approx  -4\pi n d^3 \left(\frac{\sin{x}}{x^3} - \frac{\cos{x}}{x^2}\right),
\ee
where $x=kd$ and $d=a_s+a$ is  the sum of the tracer and bath
radii. Eq.~(\ref{Sks_low_final}) shows that $S_k^s$ scales as $S_k^s = \ord(\phi)$
with $\phi$.

As for the bath structure factor, it can be written as
\be
S_k = 1 + S_k^s
\ee
for the case that the tracer is identical with the bath particles. This follows directly from the definition of $S_k$
 (see eq.~\ref{Sk_def}) since all the particles in the system are
equivalent. So, the bath
structure factor scales with $\phi$ as $S_k = 1 + \ord(\phi)$. Since the
structure factors $S_k^s$ and $S_k$ enter as a product into the relation
(\ref{delzeta_mct}), to the leading order in $\phi$, the expression
(\ref{Sks_low_final}) for $S_k^s$ together with the $0$-the order
approximation 
\be\label{Sk_low}
S_k^s \approx 1
\ee
can be used.

For the time-dependent density fluctuations $\phisk(t)$, $\phik(t)$, the
$0$-the order terms in $\phi$ are non-zero and can be obtained by neglecting the memory
integrals in eqs.~(\ref{memeq}),(\ref{memeq_phiq}) and setting $S_k=1$ in eq.~(\ref{tauq}): 
\ba\label{phisk_fex_low}
\phisk(t) & \approx & e^{D_s(-k^2 + i\, {\bf k}\cdot\bF_{ex})\, t}\, ,\\
\label{phi_low}
\phik(t) & \approx & e^{-D_0\, k^2 \, t}.
\ea
So, to the leading order in $\phi$ the higher-order corrections to
$\phisk(t)$, $\phik(t)$ can also be neglected in eq.~(\ref{delzeta_mct}).

Note that according to the translation
theorem of the Fourier transform theory, $\phisk(t)$ in
eq.~(\ref{phisk_fex_low}) is the Fourier transform of a Gaussian, 
whose maximum position  shifts by the distance $D_s\bF_{ex}t$ from the origin with the time $t$. 
It is not surprising, since we made the approximation of completely neglecting the effect of the
bath particles, i.~e.~considering a single tracer particle in a solvent,
sedimenting under the external force $\bF_{ex}$  so that it's mean position at time $t$ is given by $D_s\bF_{ex} t$. 

With the approximations (\ref{phisk_fex_low}), (\ref{phi_low}), the time
integral in eq.~(\ref{delzeta_mct}) can be performed with the result
\ba\nonumber
\int_0^\infty dt\, \phisk(t)\phik(t) = \frac{1}{(D_0+D_s)k^2 - i\,D_s{\bf
    k}\cdot\bF_{ex}}\quad\quad \\ \label{t_int_dzeta}
= \frac{(D_0+D_s)k^2 + i\, D_s{\bf k}\cdot\bF_{ex}}{(D_0+D_s)^2k^4 +
  D_s^2({\bf k}\cdot\bF_{ex})^2} \quad\quad
\ea
We notice that the imaginary part in the above expression does not contribute
to the ${\bf k}$-integral in (\ref{delzeta_mct}), since it is antisymmetric in ${\bf k}$, so that
the result for $\Delta\zeta_s$ is real.


We choose now the $k_z$-axis in the direction of the external force $\bF_{ex}$
and change to  spherical coordinates. Eqs.~(\ref{delzeta_mct}), (\ref{t_int_dzeta}) then yield
 $\Delta\zeta_s$ in the form
\be\label{dzeta_fex}
\Delta\zeta_s = \frac{1}{3}\,\varphi\,\frac{(1+\alpha)^3}{D_0 + D_s}\,f(\beta)\, ,
\ee
with the dimensionless friction increment function
\be\label{f_beta}
f(\beta) = \frac{18}{\pi}\int_0^\infty dx\,\frac{1}{\beta^3}\left(\frac{\sin x}{x} -\cos x\right)^2  
              \left(\beta -  x\arctan\frac{\beta}{x} \right),
\ee
and  the dimensionless external force parameter
\be
\beta = \frac{D_s (a+a_s) F_{ex}}{D_0+D_s}\,.
\ee
After employing the Stokes-Einstein relation (\ref{st_ein})  for $D_0$, $D_s$
and reintroducing the factor $k_B T$ the parameter $\beta$ can be  expressed as
\be\label{dmless_fex}
\beta = \frac{F_{ex} a}{k_B T}
\ee
and has thus the physical meaning of the work done by the external force over
the distance of the bath particle radius in units of $k_B T$.

\begin{figure}
\includegraphics[scale=0.7]{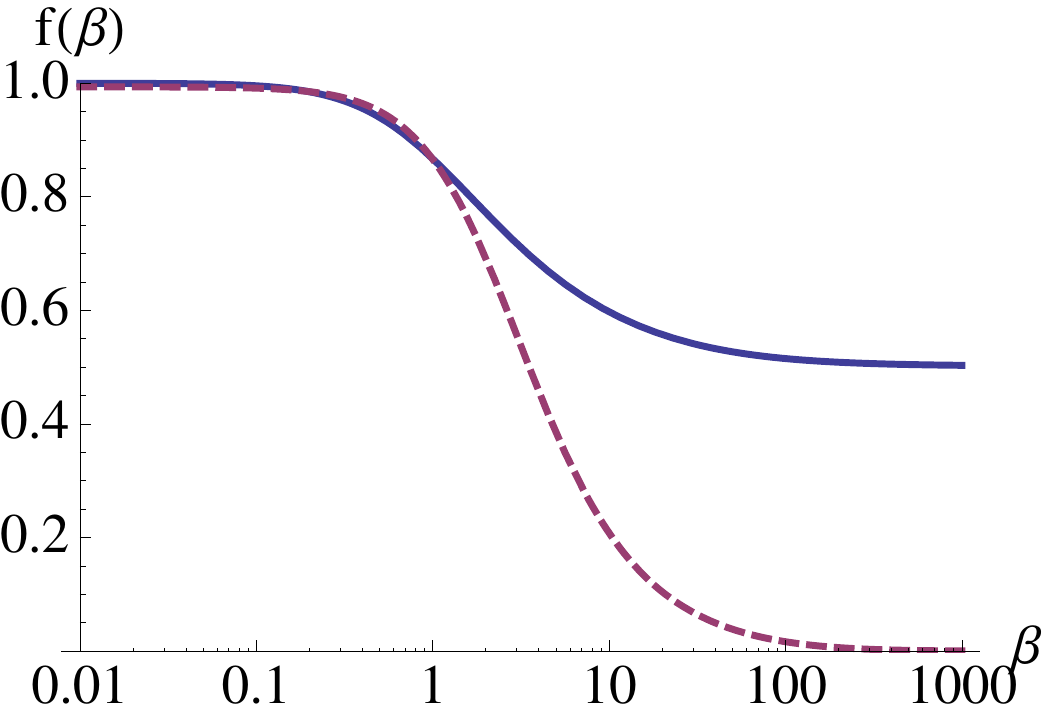}
\caption{\small\label{del_zeta_low_fig} Low density results for the dimensionless
  friction coefficient increment as function of the dimensionless external
  force parameter $\beta = \Fex a / (k_BT)$. Continuous line: exact result from Ref.~\cite{Brady05}. Dashed line: our MCT result.}
\end{figure}

We compare now our result (\ref{dzeta_fex}) for $\Delta\zeta_s$ with the exact low-density result
from Ref.~\cite{Brady05}, where the two-particle Smoluchowski equation was
solved for arbitrary external force values. The differences appear in
the numerical prefactor ($1/2$ in the exact theory instead of $1/3$ in our
calculation) and the dimensionless
friction coefficient increment function (see Fig.~\ref{del_zeta_low_fig}).
Despite these differences, the expression  (\ref{dmless_fex})  for the dimensionless external force
parameter and the scaling of the friction coefficient increment with
the system parameters $\varphi$ and $\alpha$, agree with the exact theory. 

From Fig.~\ref{del_zeta_low_fig} we see that  for values of
$\beta\lsim 2$, where the initial decay from the linear response plateau 
occurs, the MCT result (dashed line)  agrees well with the exact low-density
calculation (continuous line), whereas for values of $\beta\sim 10$ the decrease of the tracer friction coefficient with $F_{ex}$ 
is strongly overestimated by the present version of MCT.  For $\beta\rightarrow\infty$
the exact theory predicts a second plateau value of $1/2$ for $f(\beta)$,
which is zero in our MCT calculation.

Note that the approximation made to obtain the
expression (\ref{phisk_fex_low}) was to completely neglect the memory integral, and the latter also contains
the external force. Such an approximation might  not be
 valid for large $\bF_{ex}$. The large $\beta$-limit turns out to be singular
 in the low density theory \cite{Brady05} and this could also hold for our
 mode-coupling equations. This issue surely  deserves a more thorough analysis including the numerical solution of the
full time-dependent equation for the tracer density correlator.

The elementary consideration here is  just aimed to show that even
the simplest reasonable approximation (\ref{phisk_fex_low})  for the tracer correlator,  in combination with the mode-coupling
expression (\ref{corfs})  for the force-force correlator, is already
capable to explain the thinning effect. This is due to increasingly strong time oscillations
of the integrand in eq.~(\ref{delzeta_mct}), which cut off larger and larger
portions of the integrand with increasing $F_{ex}$.


\subsection[Long Time Limit of the Tracer Density Correlator]{Long time limit of the tracer density correlator}
\label{long_time_limit}

After considering the low-density dynamics in the last section,  we turn
now to the high-density ``statics''. I.e. we will assume that we are above the
critical point  where the colloidal bath forms a glass ($\varphi>\varphi_c$). The first question then concerns the existence and force 
dependence of the long time limits 
\ba
f^s_{\bf q} & = & \lim_{t\rightarrow\infty}\phi^s_{\bf q}(t)\\[1ex]
f_{\bf q} & = & \lim_{t\rightarrow\infty}\phi_{\bf q}(t)
\ea
 of the tracer and the bath density correlators. We restrict ourself here to
 the case of the equal size of the tracer and the bath particle so that 
$\alpha=1$  holds and  only the  bath volume fraction $\varphi$  remains as the  control parameter. 
The existence of a finite probe 'non-ergodicity' parameter signals that probe density fluctuations remain frozen-in, and that the 
probe can not explore the whole volume $V$. A $f^s_{\bf q}>0$ signals that the probe is localized. 

For $t\rightarrow\infty$, the full time-dependent MCT equations (\ref{memeq}), (\ref{memeq_phiq})
reduce to the ``static'' (i.~e.~time-independent) equations 
\ba\label{eq_fsq}
\frac{\omq^2\,f^s_{\bf q}}{1-f^s_{\bf q}} & = & \Msqirr(f^s,\,f),\\\label{eq_fq}
\frac{f_{\bf q}}{1-f_{\bf q}} & = & \Mqirr(f),
\ea
(the derivation of the eq.~(\ref{eq_fsq}) for the tracer straightforwardly follows the steps of the standard MCT \cite{Goetze2009}).
Equations (\ref{eq_fsq}), (\ref{eq_fq}) are  much easier to treat than the full time-dependent equations 
because no memory integrals over many time decades have to be evaluated.
Still, they contain a wave vector integration and hence represent a coupled system of equations for
 all ${\bf q}$. 

In order to be solved numerically, the long time limit equations
(\ref{eq_fsq}), (\ref{eq_fq}) have to be rewritten as fixed point equations:
\ba\label{fq_iter_eq}
\fq & = & \frac{\Mqirr}{1+\Mqirr},\\
\label{fsq_iter_eq}
\fsq & = & \frac{\Msqirr}{\omq^2+\Msqirr}.
\ea
Starting with the initial values $\fq=1$, $\fsq=1$ for all ${\bf q}$, the
right-hand side of eqs.~(\ref{fsq_iter_eq}), (\ref{fq_iter_eq})
is evaluated and then used  as input for the next iteration, until the desired precision is reached .

The spatial {\em symmetry}
properties of the tracer and bath correlators dictate the choice of 
the discretization scheme for $\fsq$ and $\fq$ in the $q$-space. 
Since the bath correlator is not affected by the external force, as was discussed in Section \ref{bulk_dyn}, $f_q$ retains
its spherical symmetry so that it can be taken as input for the equation
(\ref{fsq_iter_eq}) from the standard isotropic MCT  calculation. An equidistant
$q$-grid is chosen here with $100$ $q$-points for $qa \in (0,\,20)$
and the so-called ``Bengtzelius trick'' (see \cite{Bengtzelius1984} for details)
allows to reduce the computational complexity during
the calculation of the memory integral.

As for $f^s_{\bf q}$, the external force introduces a preferred direction in space and thus breaks the spherical symmetry. 
This means that in contrast to the isotropic calculation in the absence of $\bF_{ex}$, one explicitly has to resolve the angular structure of 
$f^s_{\bf q}$.
One symmetry  still remains, namely the rotational symmetry around $\bF_{ex}$. Thus, it is  enough to 
specify the magnitude $q$ and the angle $\theta$ between the vectors ${\bf q}$
and $\bF_{ex}$  in order to determine $f^s_{\bf q}$ uniquely. 

We thus introduce spherical coordinates with the $z$-axis pointing in $\bFex$-direction and assume  the wave vector ${\bf q}$ to
lie in the $x$-$z$ plane.  
The property $f^s_{- \bf q} = f^{s\,*}_{\bf q}$ (which we assume to be
valid, see discussion in Section \ref{real_obs_sec})  allows us to confine the
angle $\theta$ to the interval $(0,\frac{\pi}{2})$.
Note however that despite this quasi-2d nature of the problem, the integration
in the memory kernel includes all the wave vectors and cannot be reduced to a two-dimensional 
one. So, the polar angle $\phi$ also has to be included there, too. 

Since the angles $\theta$  enter in eqs.~(\ref{fsq_iter_eq}) only via scalar
products of  vectors ${\bf q}$, ${\bf p}$ and $\bFex$ with each other  (see relations (\ref{omq_Fex}),
(\ref{memfq_fex}) for $\omq$ and $\Msqirr$), 
i.~e.~only as  $\cos\theta$ or $\sin\theta$,  it appears reasonable to change
the variable from $\theta$ to $y=\cos\theta$  and  employ the relation $\int_0^\pi d\theta\, \sin\theta\, f(\theta) =
\int_0^1 dy\, f(y)$ in calculating the wave vector integral entering $\Msqirr$.

Since the  $\fq$-s enter the tracer equations (\ref{fsq_iter_eq}) as  input,
the same equidistant $q$-grid is chosen for $\fsq$ as for $\fq$. The $y$- and $\phi$-grids are also chosen
equidistant with  $30$ points for $ y \in (0,1)$ and $20$ points for
$\phi \in (0,2\pi)$.

Finally, as the equilibrium structural input, the  Percus-Yevick structure factor
\cite{Hansen1986,Percus1958} is used.

\begin{figure*}
\begin{tabular}{c}
\includegraphics[scale=0.41]{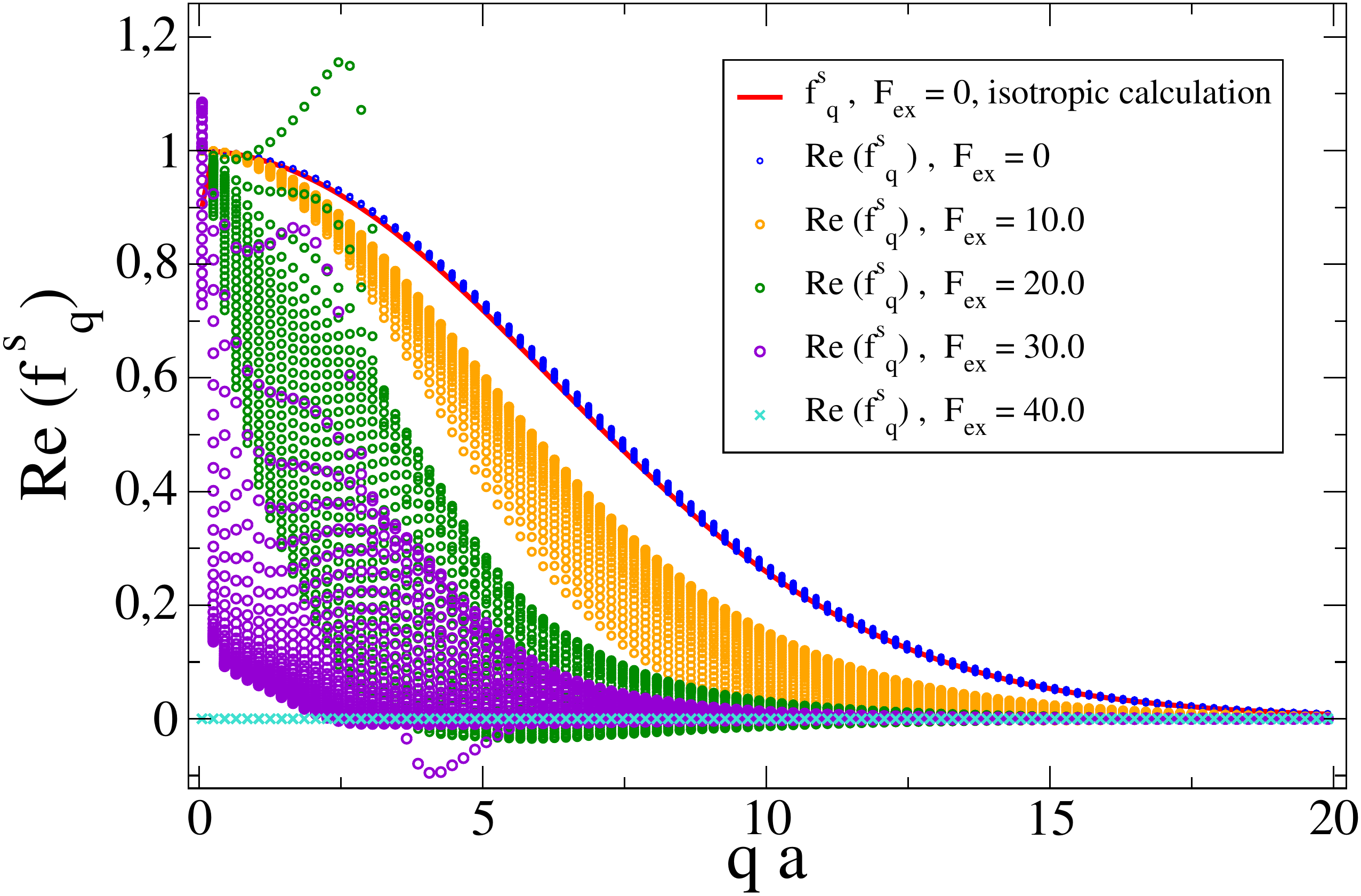}\\[1ex]
\includegraphics[scale=0.41]{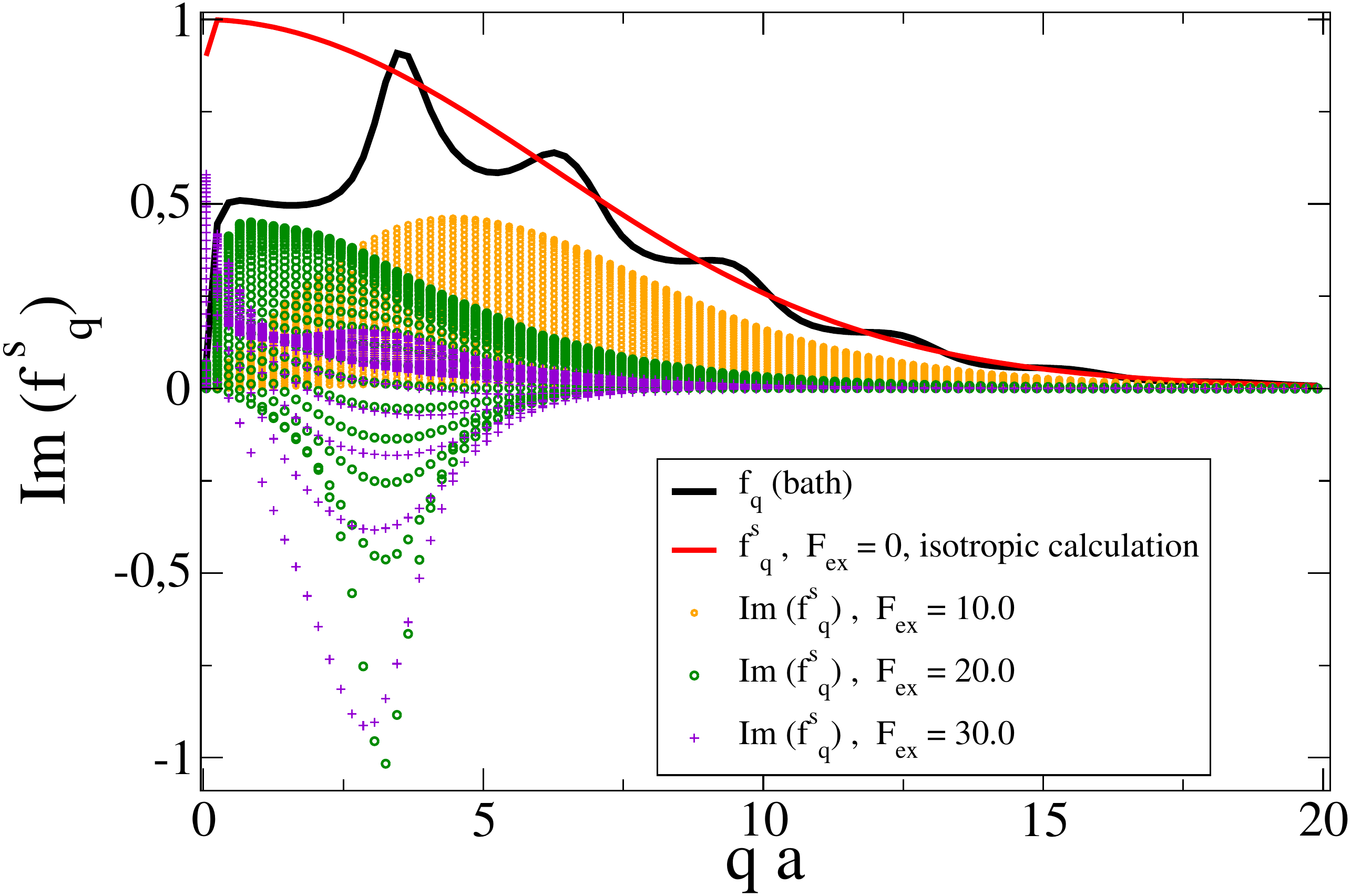}
\end{tabular}
\caption{\small\label{fqs_Fex_fig} (Color online). Long time limit of the tracer density
  correlator (upper panel  shows the real and lower panel the imaginary parts)
  as function of the wave vector magnitude $q$ for different angles $\theta$ (see text) and different $F_{ex}$. 
The bath volume fraction is $\varphi=0.52$. The two $f^s_q$ at $\Fex=0$ in the upper panel show that the used angle-discretization is fine enough 
to recover the known isotropic result. In the lower panel the collective $f_q$ is included as reference. The rapid variations in all $f_q$ below $qa<0.5$ are 
discretization artifacts which do not affect the results for larger $q$.}
\end{figure*}

The results of our numerical calculations for $\fsq$  are shown in 
Fig.~\ref{fqs_Fex_fig}. The value of the control parameter $\varphi$ was choose
to be $0.52$, i.~e.~closely above the glass transition point $\phi_c \approx 0.516$ of quiescent MCT. 
The real and the imaginary  parts of $f^s_{\bf q}$ are plotted as function of $q$ and 
the different curves for the same force value correspond to the different
values of angle $\theta$.  $F_{ex}$ is given  in the dimensionless units  of $k_B T/a$ (see eq.~\ref{dmless_fex}). The
 magnitude of the wave vector $q$ is given in units of $1/a$.

For $F_{ex}=0$, no angle dependence is observed and the standard isotropic MCT result is recovered.
As $F_{ex}$ increases from $0$ and $30$,  the splitting of $\fsq$-s with $\theta$
becomes more and more pronounced, revealing the 
spatial anisotropy, which increases with the external force. 

For $F_{ex}>40$, $\fsq$ appears to drop to zero for all ${\bf q}$. 
This is an indication of the {\em bifurcation} transition in the equations (\ref{eq_fsq}): there is a certain {\em critical} 
force $F^c_{ex}$, 
above which only the (always existing, trivial)  zero solution of (\ref{eq_fsq}) becomes stable, whereas below $F^c_{ex}$ the 
numerically found non-zero solution is stable.

Unfortunately, the numerics becomes unstable for $30 < \Fex < 40$, so that we
cannot approach the region of the critical force value arbitrarily closely. But the observed
trends in the $F_{ex}$-dependence of the curves suggest that the transition is continuous 
(``of type A'' in the MCT classification). I.~e.~for every ${\bf q}$, $f^s_{\bf q}$ goes
down to zero continuously with increasing $F_{ex}$. At the bifurcation point,
$f^s_{\bf q}$ becomes zero so that the two solution branches of eq.~(\ref{eq_fsq}) coalesce.

The physical meaning of the bifurcation transition is  that the cage surrounding the probe becomes weaker with increasing
external force until the tracer gets ``pulled free'' out of the cage and {\em depins} from the glassy matrix. 
In \cite{Gazuz2009}, we showed the results for the quantity $\fsr$, which is the Fourier backtransform of $\fsq$  and has
a very clear physical meaning; see eq.~(\ref{phisr_def}) and discussion after it.  It  can be related to the ``shape'' of the cage.
The results for $\fsr$, discussed in \cite{Gazuz2009},  were obtained numerically  from the $\fsq$, which seem not so easy to interpret, at first glance. It is still possible, however, to relate the observed features
in the behaviour of $\fsr$ (see Fig.~2 of Ref.~\cite{Gazuz2009}) to the features of the $\fsq$-curves (Fig.~\ref{fqs_Fex_fig}).

First, we see that for $\Fex=10$, the real parts of $\fsq(q)$ are
monotonously decaying and the imaginary parts  have one maximum for
all values of $\theta$. Furthermore, the shape of the curves resembles the one
obtained from the $\theta$-independent real function $\fsq(\Fex=0)$ by
multiplying by the prefactor $\exp( i\,{\bf q}\cdot \delta{\bf  r} ) = \cos
(q\,\delta r\cos\theta) + i\sin( q\,\delta r\cos\theta)$:
\be\label{fsq_small_fex}
\fsq(\Fex) \approx \fsq(\Fex=0)\,\exp( i\,{\bf q}\cdot \delta{\bf  r}(\Fex)),
\ee
where we assume  $\delta{\bf r} || \bFex$. Due to the translation theorem of the Fourier transform
theory, this prefactor corresponds to the shifting of the
$\fsr(\Fex=0)$-distribution by a vector $\delta {\bf r}\sim \bFex$. This is in
accordance with our observations for the behaviour of $\fsr$.
For higher values of $\Fex$, for some values of $\theta$ the real parts of the curves $\fsq(q)$ become
non-monotonous, whereas the imaginary parts
exhibit a maximum and a minimum. This complicated behaviour causes
subtle features in the shape of the $\fsr$-distribution discussed in \cite{Gazuz2009} (anisotropy with respect to the maximum, ``dent''). 

Finally, we  discuss some peculiar features in the behaviour of the
imaginary part of $\fsq$. 
For the direction perpendicular to $\bFex$, the imaginary
part of  $f^s_{\bf q}$ should {\em vanish}. This is the consequence  of the relation
$f^s_{\bf -q} = f^{s\,*}_{\bf q}$ and the rotational symmetry  around
$\bF_{ex}$, which implies     $f^{s\,*}_{{\bf q}^\perp}=f^s_{{\bf q}^\perp}$ for ${\bf q}^\perp \cdot {\bf F}_{\rm ex}=0$,
so that
\be
\Im \{\fsq\} = 0 \quad\quad \mbox{for}\quad {\bf q}\bot\bF_{ex} 
\ee
should apply.

For $F_{ex}=10$, we indeed observe this behaviour
(which is also in agreement with our approximation (\ref{fsq_small_fex})), 
whereas for higher force values, a strong dip is observed for $\Im \{\fsq\}
(q)$ at  $qa\approx 3$, for the $\theta$-values closest to $\pi/2$, whereas at
$q=0$,  $\Im \{\fsq\}(q)$  seems to diverge for all $\theta$.
Since  $\Im \{f^s_{\bf  q}\}$ is antisymmetric in $\bq$, this behaviour
signalizes the  discontinuity of the function $\Im \{\fsq\} (\bq)$   at the plane $\bq\bot\bFex$.

It is easy to see that the origin of this discontinuity lies in the presence
of the term $\omq^2$, which enters the long time limit eq.~(\ref{fsq_iter_eq})
under the denominator on the right hand side. 
 The term $\omq =  D_s\, ( q^2 -i\,{\bf  q}\cdot\bF_{ex})$ itself obviously has the property $\Im \{\omq\} = 0$  for
${\bf q}\bot\bF_{ex}$, so the same should be valid for its inverse. The
inverse however, is divergent at the point $\bq=0$, so that $\Im\{ 1/\omq\} = 0$ applies
everywhere on the plane ${\bf q}\bot\bF_{ex}$  except for the singular
point ${\bf q}=0$. 

The discussed behaviour of the term $1/\omq$ is relevant for $\fsq$, since if
one approaches the bifurcation transition, the 
function $\fsq(\bq)$ becomes confined to a narrow  region around the origin in the $\bq$-space and the
same applies also for the memory function $\Msqirr$. So, as a very rough first
guess, $\Msqirr$ can be neglected in the denominator of
eq.~(\ref{fsq_iter_eq}), as a ``small'' quantity.
As a next approximation,  one can use an  $\bq$-independent ansatz for
$\Msqirr$ in (\ref{fsq_iter_eq}). In this way it is already possible to obtain the structure of the $\bq$-dependence very similar
to that observed for  $\fsq$  in our numerical solution of
eq.~(\ref{fsq_iter_eq}) at high force values.  

We believe that the qualitative discussion of the properties of $\fsq$
given here, already contributes to a deeper understanding of our numerical
results. The origin of the unphysical results for $f^s_{\bf q}$ at larger
forces remains to be clarified.  It should be possible to extract also quantitative results for the
limiting cases $\Fex\rightarrow 0$ and $\Fex\rightarrow \Fex^c$ 
by either performing expansions in $\Fex$ (for
 $\Fex\rightarrow 0$) in eq.~(\ref{fsq_iter_eq}) or an appropriate asymptotic expansion
assuming the smallness of $\fsq$ (for $\Fex\rightarrow \Fex^c$). This will be a
subject of  future work. For the case of schematic models, the latter has recently been achieved \cite{Gnann2012}.


\section{Schematic Models}
\label{schem_models}

The bifurcation scenario we deduced from the wavevector dependent equations of motion is continuous. At the critical force $\Fex^c$, the long time limit $f^s_{\bf q}$ vanishes,  as has been found in type A transitions within quiescent MCT  \cite{Goetze2009}. The universal properties close to quiescent type A transitions could be analyzed in schematic models, because all wavevectors are coupled strongly at the bifurcation. Here we show that the simplest schematic models containing complex correlators, which can be derived from the full theory under external force, again describe a continuous delocalization transition at a critical force. They can thus  be used to gain insight into the more universal phenomena close to the bifurcation.

\subsection{Construction of the models}

Simplified schematic models for tracer density fluctuations can be constructed by considering only two wave vectors ${\bf k}, -{\bf k}$ 
with ${\bf k}\,||\,\bF_{ex}$ and neglecting the contribution of all the other wave
vectors to the memory function. Due to the  property   $\phismk = \phi^{s\,*}_{\bf k}$, the system of
equations for $\phisk$, $\phismk$, obtained in this way, reduces to the following single equation for $\phisk$:
\ba\label{phisk_sch}
\partial_t\, \phisk(t)   =  -D_s(k^2 - i k F_{ex})\,\phisk(t)\quad\quad\quad\quad\quad\quad\quad\quad \\ \nonumber
 -  \int_0^t dt'\, v_{\bf k}k^2\,\phisk(t-t')^*\,\phi_{2{\bf k}}(t-t') \, 
          \partial_{t'}\phisk(t') \,,
\ea
where $v_{\bf k} = 4 {S_k^s}^{\,2} / N S_{2k}$.
We see that the force dependence of the memory function drops in eq.~(\ref{phisk_sch}).
Further simplification can be achieved if we introduce a scale for lengths so
that   $k=1$,  rescale time so that $D_s=1$ and model the behaviour of the bath correlator 
$\phi_{k=1}\equiv \phi$ with the standard F12-model \cite{Goetze91}. For simplicity we set its short time coefficient equal to the tracer's. 

We finally arrive at
\ba
\!\!\!\!\!\!\!\! \partial_t\, \phi^s(t)  & = & -\om\,\phi^s(t) - \int_0^t dt'\,m^s(t-t') \, 
 \partial_{t'}\phi^s(t'), \label{fex_sjgr}\\
\!\!\!\!\!\!\!\! \partial_t\, \phi(t) &   = &  - \phi(t) - \int_0^t dt'\, m(t-t')\, \partial_{t'}\phi(t'),
\ea
with the complex frequency
\be\label{om_fex}
\om  =  1- i\,F_{ex}
\ee
and the memory functions
\ba
m^s(t) & = & v_s\,\phi^{s*}(t)\phi(t)\\
m(t) & = & v_1\,\phi(t) + v_2\,\phi^2(t).
\ea

This schematic model will be called the {\em $F_{ex}$-Sj\"ogren model}. It
extends the original Sj\"ogren model (proposed in Ref.~\cite{Sjoegren86}) for the tagged particle dynamics in a
host fluid for the case of external driving with the force $\Fex$.  
From the described rescalings, it is obvious that in the schematic model $\Fex$ is measured in thermal energy divided by the length 
used to rescale the wavevector in Eq.~(\ref{phisk_sch}). 

A modification of the model (\ref{fex_sjgr})  is of interest, namely the one where $\phi(t)\equiv 1$ is set:
\be\label{fex_f1}
\partial_t\, \phi^s(t)   =  -\om\,\phi^s(t) 
 - \int_0^t dt'\, v_s \,\phi^{s*}(t-t') \, 
 \partial_{t'}\phi^s(t')\ .
\ee
We call this model the {\em $F_{ex}$-F1 model}. It extends the well-known $F1$-model  \cite{Goetze91}
(corresponding to the case $F_{ex}=0$) which was introduced in connection with
the mode-coupling theory for the Lorentz model. The latter describes a tagged
particle in an array of immobile randomly distributed scatterers. As we shall see in Sec.~\ref{fric_coef_sec_1}, 
the $z=0$ value for the Laplace transform of our $F_{ex}$-F1 model is available in exact form.

\subsection{The phase diagrams}\label{sch_ph_dagr_sec}

\begin{figure}
\includegraphics[scale=0.7]{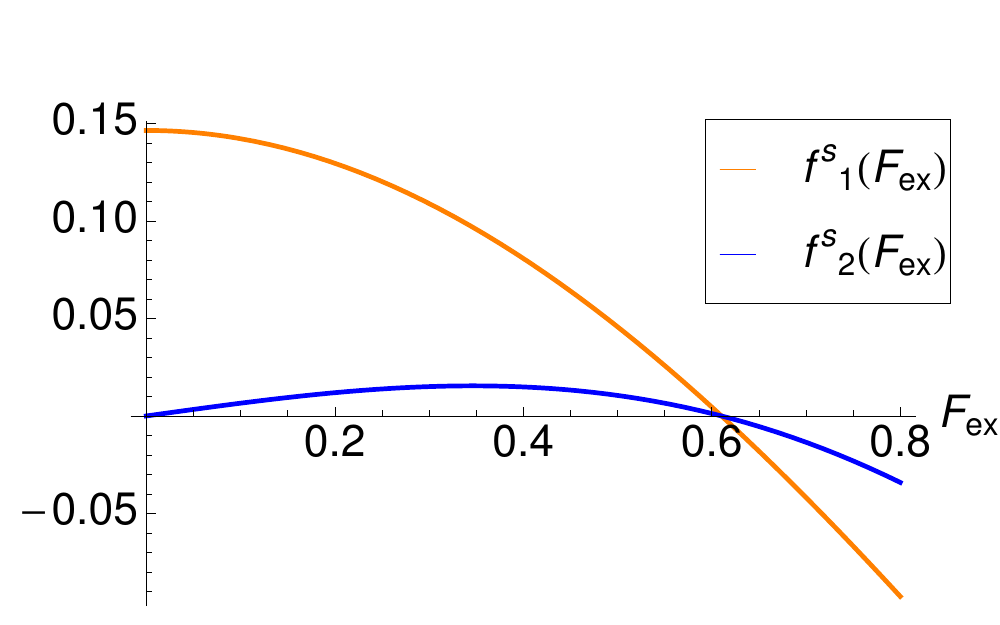}
\caption{\small\label{phis_vs_fex} Real and imaginary parts of the long time limit of the tracer density correlator 
as functions of the external force ($v_2=2.0$, $v_s=4.0$).}
\end{figure}

In this section we consider the long time limit $f^s=f^s_1+if^s_2$ of the
correlator $\phi^s(t)$. For our schematic models, the condition (\ref{eq_fsq}) reduces to
\be\label{sch_ltlim_eq}
\frac{\om\,f^s}{1-f^s} = m^s(f^s, f),
\ee
Considering the real and imaginary of eq.~(\ref{sch_ltlim_eq}) separately, we obtain the system of equations
\ba\label{eq_fs1}
F_1(f^s_1,\,f^s_2) =0\\\label{eq_fs2}
F_2(f^s_1,\,f^s_2) =0
\ea
for $f^s_1$, $f^s_2$, where
\ba\label{eq_F1}
F_1 & = & f^s_1 + F_{ex} f^s_2 - v_s f \,(f^s_1 - {f^s_1}^2 - {f^s_2}^2 )\\
\label{eq_F2}
F_2 & = & f^s_2 - F_{ex} f^s_1 + v_s f f^s_2.
\ea
Eq.~(\ref{eq_fs2}) yields the linear relationship 
\be\label{fs2}
f^s_2   =  \alpha\,f^s_1
\ee
 between  $f^s_2$ and $f^s_1$ with
\be\label{alpha}
\alpha  =  \frac{F_{ex}}{1 + v_s\,f}\, ,
\ee
which can be inserted into eq.~(\ref{eq_fs1}) to obtain a quadratic equation for $f^s_1$. Its first trivial solution is $f^s_1=0$. 
The second solution is:
\be\label{fs_ltlimit}
f^s_1 = \frac{ - \alpha\,F_{ex} + v_s\,f - 1}{ v_s\,f\,(1 + \alpha^2)}\;.
\ee

In the absence of the external force we have $\alpha=0$, so $f^s_2=0$. I.~e.
the tracer correlator is real and the result of the original Sj\"ogren model is recovered:
\be\label{sjgr_ltlimit}
f^s_1 = 1-\frac{1}{v_s\,f}\,.
\ee
There is a {\em bifurcation}  at $v_s^c=1/f$: for $v_s< v_s^c$, the long time limit of $\phi_s(t)$ is zero 
(since it cannot be negative) and for $v_s>v_s^c$ it has 
a non-zero value which is given by eq.~(\ref{sjgr_ltlimit}). This
consideration is valid for $f > 0$, i.~e., if the bath is in a glassy state. If $f = 0$,  then $f^s_1 = 0$  always holds.

Now, let us assume that for $F_{ex}=0$,  the tracer is in an arrested state: $f^s_1>0$. Can this state be ``molten'' by the 
external force field ? We  look at the behaviour of $f^s_1$ as a function of the external force for fixed values of $f$ and $v_s$.
Eqs.~(\ref{alpha}),(\ref{fs_ltlimit}) yield:
\be\label{fs1_Fex}
f^s_1 = \frac{(1+v_s\,f)((v_s\,f)^2 - 1 - F_{ex}^2)}{v_s\,f\,((1+v_s\,f)^2 + F_{ex}^2)}
\ee

In Fig.~\ref{phis_vs_fex}, $f^s_1$ and $f^s_2$ are plotted as function of
$\Fex$ for fixed parameter values $v_2=2.0$,  $v_s=4$,  .
We see that our schematic model  exhibits the {\em critical} force at which the tracer becomes delocalized, 
i.~e.~where the depinning transition occurs.

\begin{figure}
\begin{tabular}{c}
\includegraphics[scale=0.35]{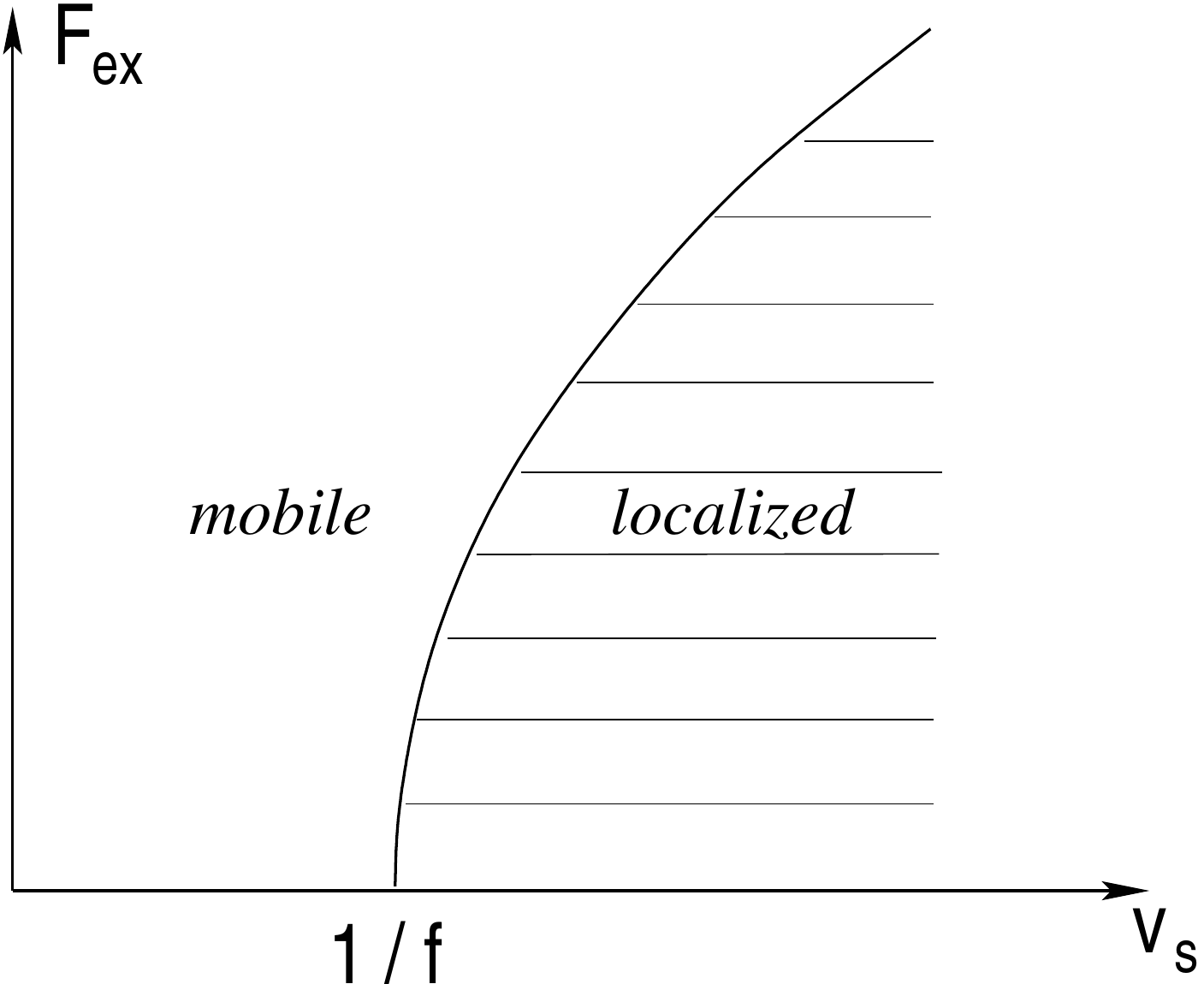}\\[6ex]
\includegraphics[scale=0.35]{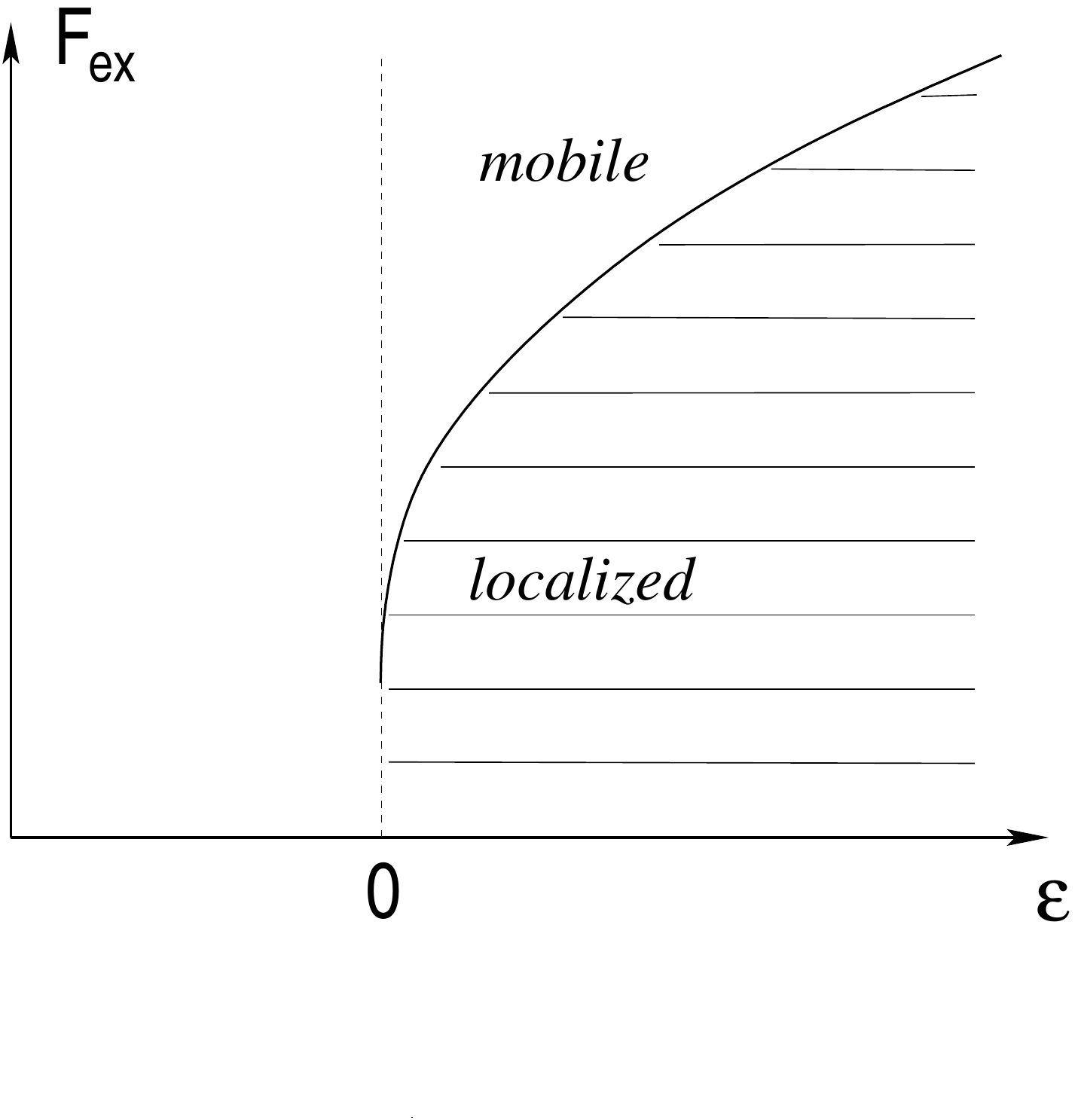}
\end{tabular}
\caption{\small\label{ph_diagr_schem} Phase diagram sketches of the schematic model for fixed $f>f^c$ 
(upper panel) and fixed $v_s$ (lower panel).}
\end{figure}

From eq.~(\ref{fs1_Fex}), we obtain 
\be\label{cr_Fex_eq}
F^c_{ex} = \sqrt{(v_s\,f)^2 - 1}
\ee
for the critical force $F^c_{ex}$, at which real and imaginary part of  $f^s$
become zero together. 
Eq.~(\ref{cr_Fex_eq}) can be visualized in form of phase diagrams, (schematically) shown
in Fig.~\ref{ph_diagr_schem}. Note that in the fixed $v_s$ diagram the horizontal axis corresponds to the variable $\eps(f)$, 
which measures the distance from the glass transition point of the bath (see Sec.~\ref{ext_force_eff} for the definition) so that
$\eps=0$ corresponds to $f=f^c$. 


The analytical results for the long time limit derived above for the
$F_{ex}$-Sj\"ogren model obviously apply for the $F_{ex}$-F1 model 
if one  sets $f=1$ in the corresponding expressions.

\subsection{Bifurcation analysis}\label{bif_an_sec}

The purely algebraic consideration in the last Section can be completed by the analysis of the bifurcation scenario
for the eq.~(\ref{sch_ltlim_eq}). 

At the critical force, the two solutions $f^s=0$ and the one given by eqs.~(\ref{fs_ltlimit}), (\ref{fs2})  coalesce.
In order to clarify the geometry of the problem, we look at the solution space of eq.~(\ref{sch_ltlim_eq}), namely the
($f^s_1$, $f^s_2$)-plane. Since the complex conjugation is involved, being  a nonanalytic operation, no use can be made of the 
complex analysis 
\footnote{\label{ftnote_model_no_star} Interestingly enough, in absence of the complex  conjugation in the memory 
function of the schematic model, no yielding behaviour  is observed.}. %

So, instead of the complex eq.~(\ref{sch_ltlim_eq}) we have to analyze the equivalent system of real equations (\ref{eq_fs1}), (\ref{eq_fs2}). 
Each of them defines a curve in the solution plane. The curves are of the order not higher then
quadratic. Generically, they intersect in two points. The  bifurcation occurs, when the two curves are just tangent to each 
other, so that there is only one intersection point identical with the
osculation point.

The normals to the curves  defined by eqs.~(\ref{eq_fs1}) and (\ref{eq_fs2}) are given by the gradient vectors
$(\partial F_1/\partial f^s_1,\, \partial F_1/\partial f^s_2)$ and $(\partial F_2/\partial f^s_1,\, \partial F_2/\partial f^s_2)$, 
respectively. They have to be {\em parallel} at the bifurcation point (since the curves have to be tangent, as discussed above) and
this is equivalent to the requirement 
\be\label{detJzero}
\det J = 0
\ee
 for the  matrix $J$ with the elements $J_{ik}=\partial F_i/\partial f^s_k $, where  $i,k \in \{1,2\}$.
From eqs.~(\ref{eq_F1}), (\ref{eq_F2}) we obtain:
\be\label{matr_j}
J = 1\!\!\mbox{I} - I
\ee
with
\be\label{matr_i}
I =\left(
\begin{array}{cc}
v_s f & -F_{ex}\\
F_{ex} & - v_s f
\end{array}\right),
\ee
and the unity matrix $1\!\!\mbox{I}$. So, condition (\ref{detJzero}) is
equivalent to the condition $\det I = 1$, which immediately yields 
the expression (\ref{cr_Fex_eq}) for the critical force.

Besides from recovering the result for the critical force from the last section, the considerations here enable us  to learn more
about the bifurcation. 
Let us look at the eigenvalues $\lambda$ of the matrix $I$. They are solutions of the characteristic equation
$\det (I-\lambda\, 1\!\!\mbox{I})=0$, which yields
\be
\lambda^2 =  v_s^2 f^2 - F_{ex}^2.
\ee
The eigenvalue $\lambda=1$ corresponds exactly to $F_{ex}=F^c_{ex}$. This
eigenvalue is not degenerated. This means, the bifurcation is of the {\em codimension} one \cite{Arnold1986}.

\subsection[Effect of the external force on the correlators]{Effect of the external force on the correlators}\label{ext_force_eff}

\begin{figure*}
\begin{tabular}{lr}
\hspace{-3ex}
\includegraphics[scale=0.27]{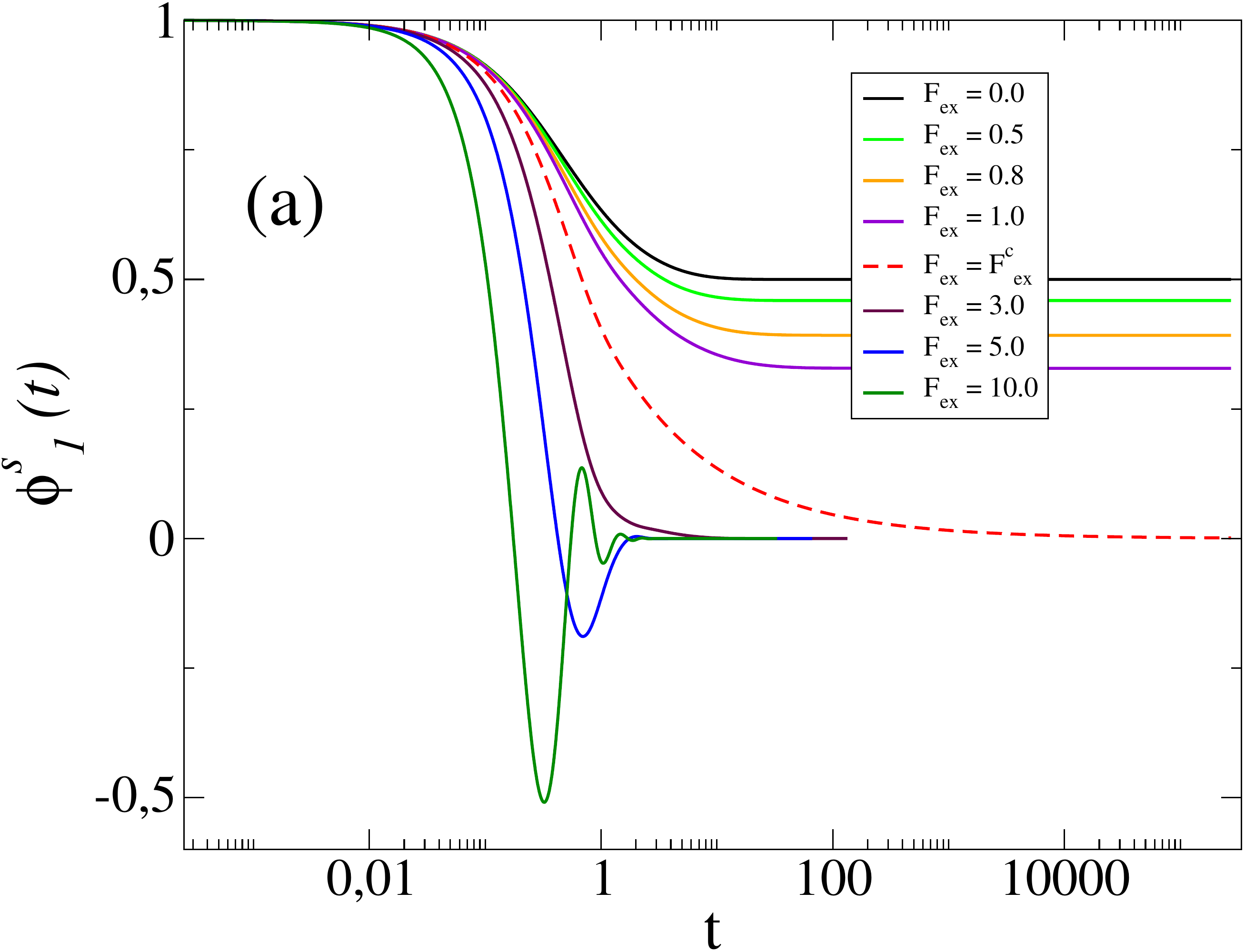}\hspace{2ex} & 
\includegraphics[scale=0.27]{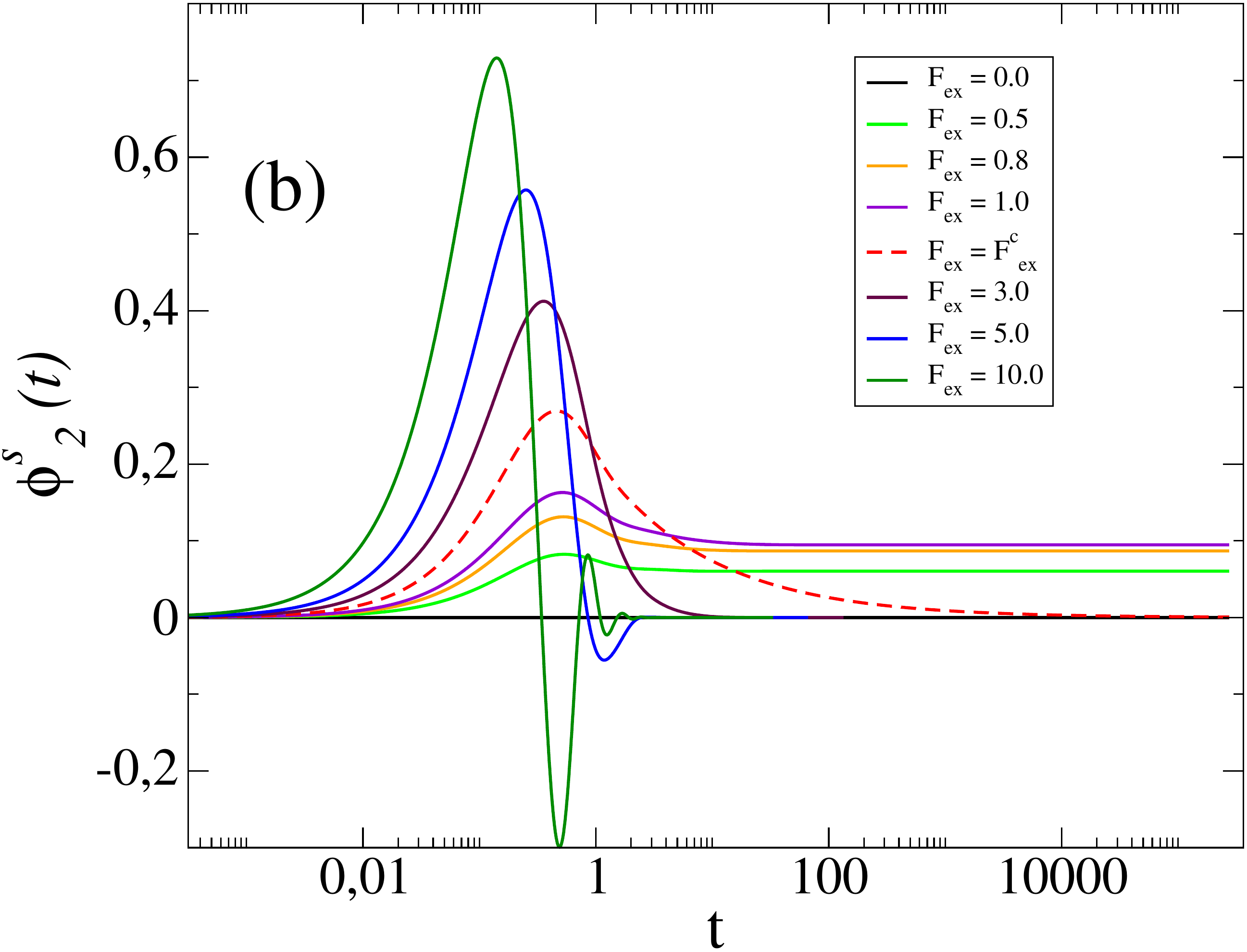}\\

\hspace{-3ex}
\includegraphics[scale=0.27]{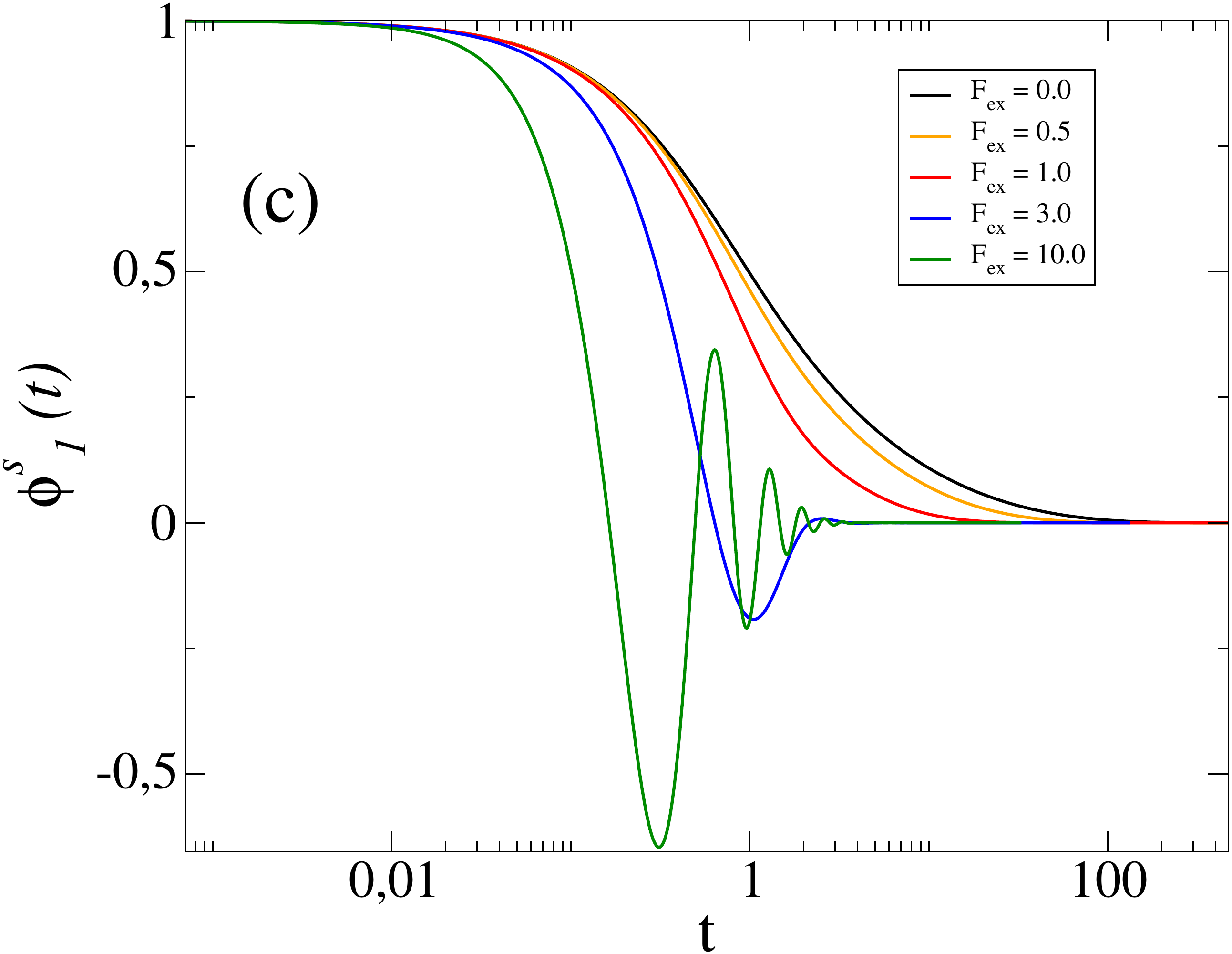}\hspace{2ex} &
\includegraphics[scale=0.27]{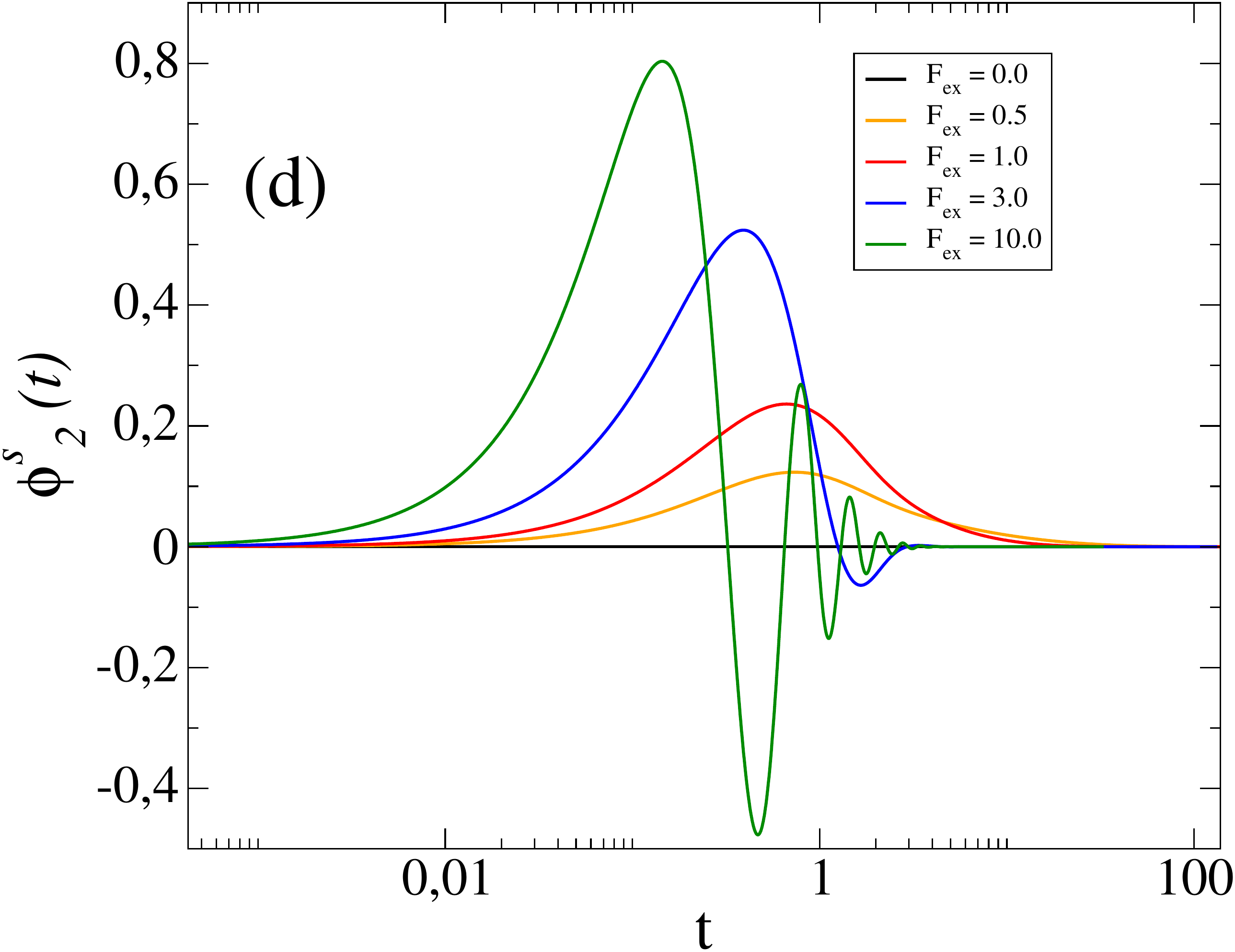}\\

\hspace{-2ex}
\includegraphics[scale=0.27]{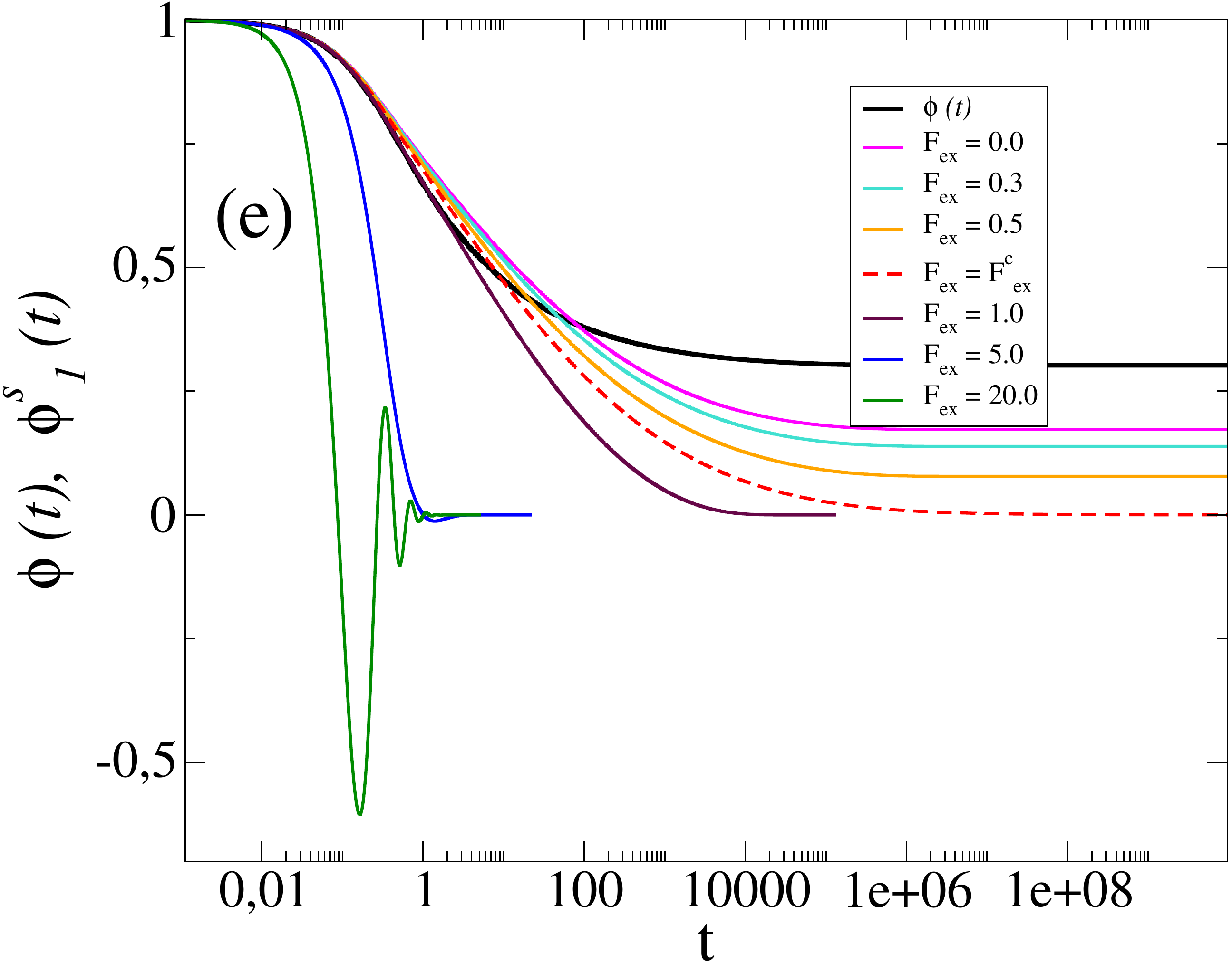}&\hspace{2ex}
\includegraphics[scale=0.27]{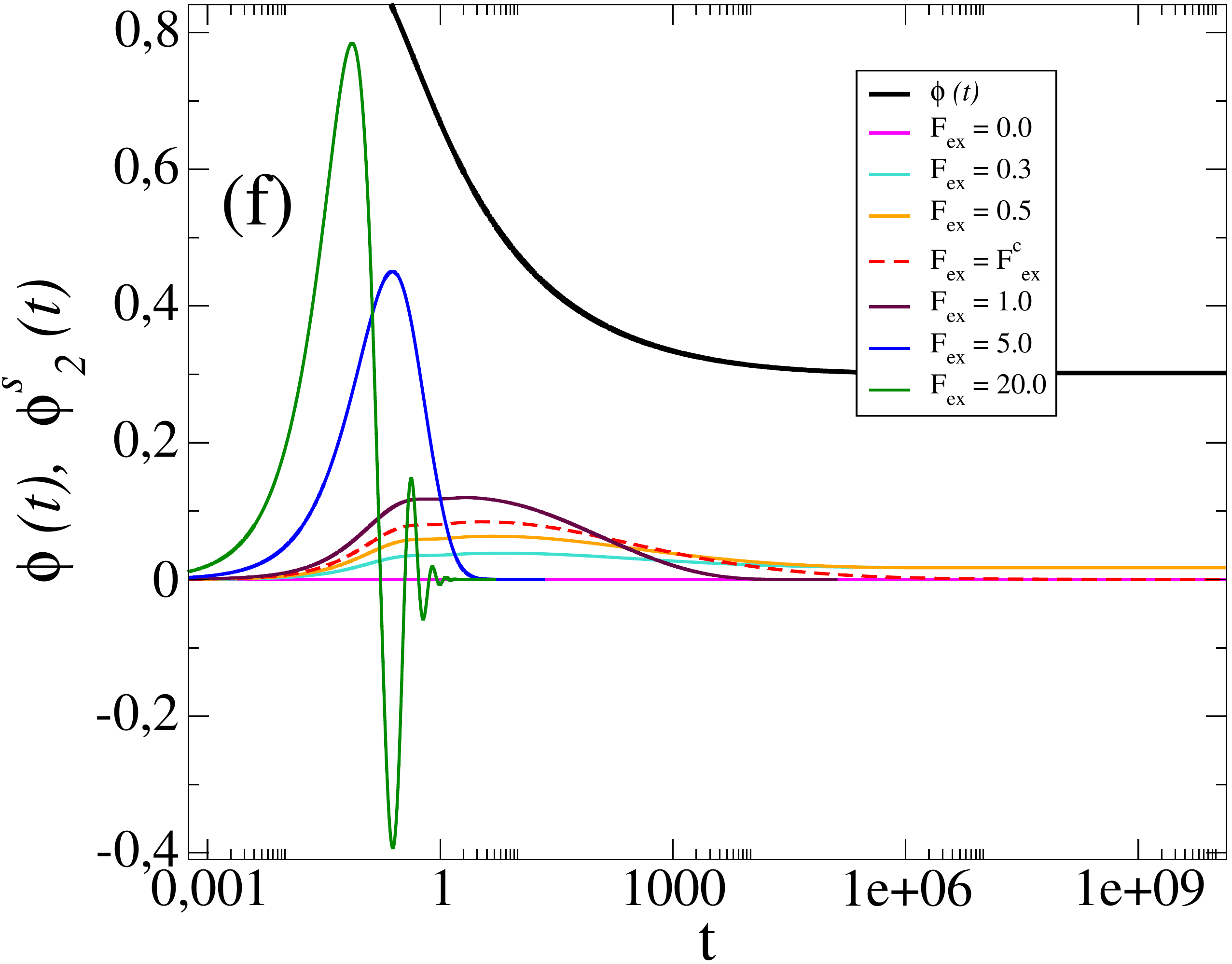}\\

\hspace{-2ex}
\includegraphics[scale=0.27]{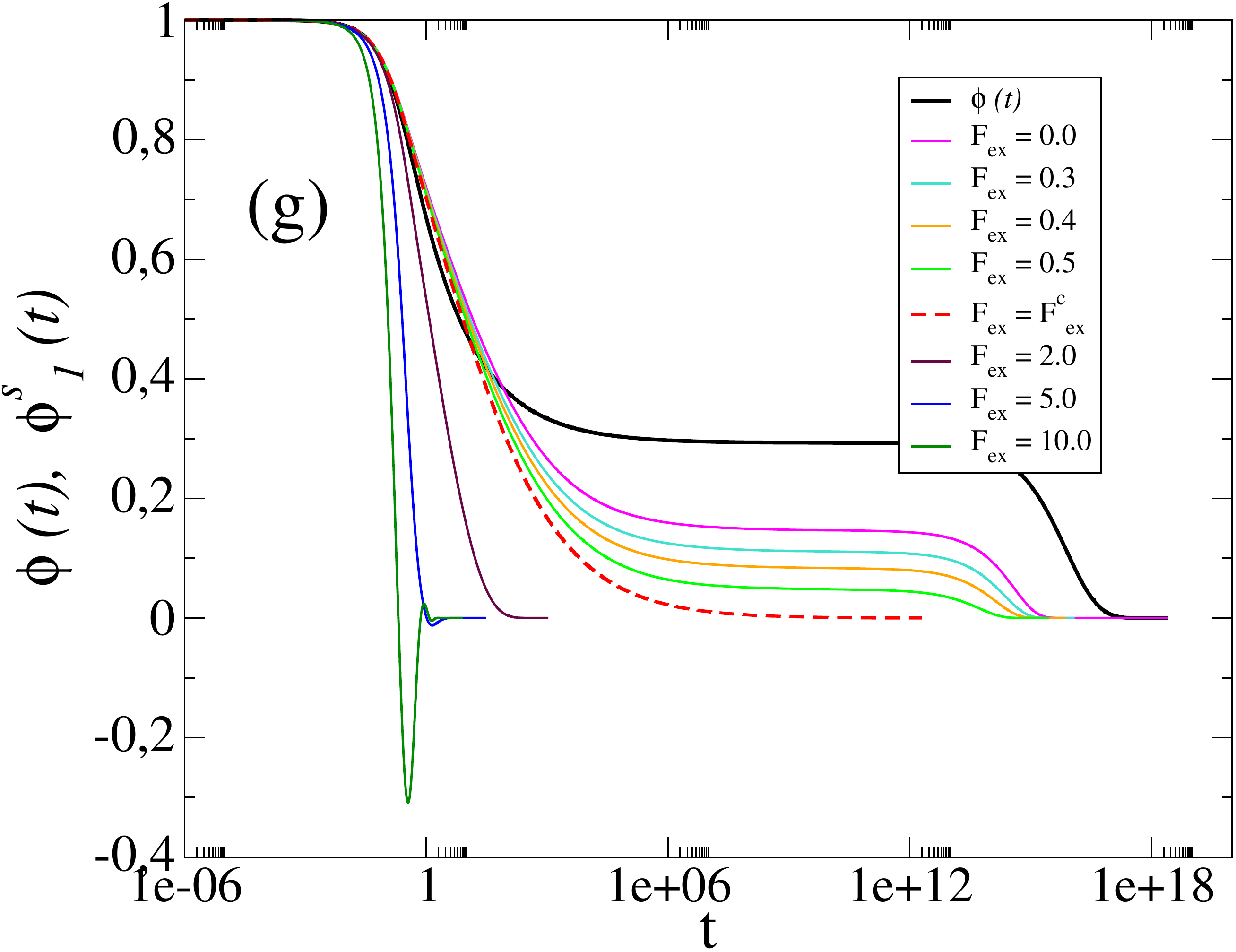} &\hspace{2ex}
\includegraphics[scale=0.27]{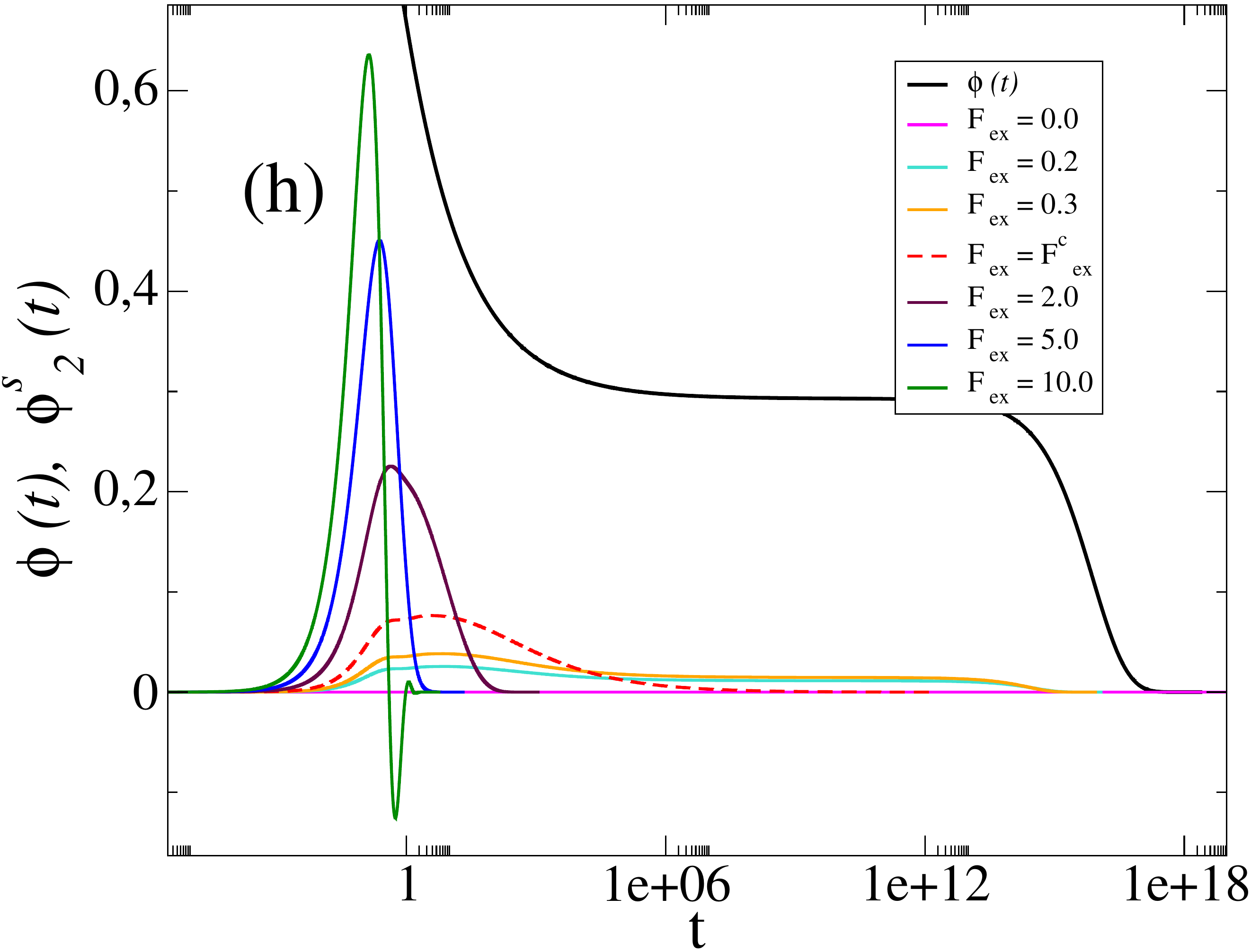}
\end{tabular}
\caption{\small
\label{fex_corr_fig} (Color online). 
Time-dependence of the probe-particle correlators from the schematic models
with increasing $F_{ex}$. The left column shows the real parts, the right column the 
imaginary parts.  Panels (a-f) correspond to glass 
states, where the bath is nonergodic; panels (g,h) are for a fluid state,
 here $F_{\rm ex}^c=F_{\rm ex}^c(\epsilon=0)$. 
Panels (a-d) are for the $\Fex$-F1 model where the bath is  completely arrested
($\phi=1$), and panels (e-h) are for the $F_{ex}$-Sj\"ogren model. Panels (c,d) are for a weakly coupled probe ($v_s<v_s^c$), 
which remains mobile in glass even at vanishing force.
 The parameter values are $v_s = 2.0$ (a,b), $v_s = 0.8$ (c, d) and
 $\eps=10^{-4}$, $v_2=2.0$, $v_s=4.0$ (e, f),  $\eps=-10^{-7}$, $v_2=2.0$, $v_s=4.0$ (g, h). }
\end{figure*}

We want to consider now the full time dependence of the correlators and look, how it is influenced by the external force.
The equations (\ref{fex_sjgr}), (\ref{fex_f1})  are solved  numerically 
using the algorithm described in \cite{Fuchs1991b} applied to real and
imaginary part individually. The results are summarized in
Fig.~\ref{fex_corr_fig} for typical values of the parameters.

The behaviour of the schematic models of the standard MCT
well known from Refs.~\cite{Sjoegren86,Goetze91,Goetze2009} will be taken as
reference where appropriate in the following. 
For the parameters of the F12 model the relation holds
\be\label{vc1_vc2_eq}
v^c_1 = v^c_2\,(\frac{2}{\sqrt{v^c_2}} -1),
\ee 
for $v^c_1$, $v^c_1$ lying on the bifurcation line, separating the liquid phase from the glass phase.
We introduce also the  parameter $\eps$ which measures the distance from the glass transition line and
 is related to $v_2=  v^c_2 + \delta v_2$ and $v_1 = v^c_1 + \delta v_1$ by
\be\label{eps_deltav1_eq}
\eps = \frac{\delta v_1\, f^c + \delta v_2\, {f^c}^2 }{1-f^c}
\ee
with $f^c = 1- \frac{1}{\sqrt{v^c_2}}$.
In the following, the parameters will be chosen such that $v_2=v^c_2$, 
so that $\delta v_2=0$. Thus, specifying the values of $v_2$ and $\eps$ completely determines the parameters
of the F12-model according to the eqs.~(\ref{eps_deltav1_eq}), (\ref{vc1_vc2_eq}).

First we consider the case, where the tracer is in the  arrested state for $\Fex=0$. For the $\Fex$-F1 model this means
 $v_s > 1$, and for the $F_{ex}$-Sj\"ogren model $\eps>0$ and $v_s > 1/f$.
The corresponding plots are shown on panels (a,b) (for the  $F_{ex}$-F1 model)
and (e,f)  (for the $F_{ex}$-Sj\"ogren model) of Fig.~\ref{fex_corr_fig}. We
can see that  the behaviour of the  long time limit of
the tracer correlator is in accordance with the results of
Sect.~\ref{sch_ph_dagr_sec} (see Fig.~\ref{phis_vs_fex}). For the real part, the long time limit goes down
to zero monotonously with  increasing $\Fex$, whereas for the imaginary part, the long time limit
first increases from the zero value at $\Fex=0$ and then goes down to zero,
until the critical force value $\Fex^c$ is reached (see the dashed thick red line in the
plots). For $\Fex>\Fex^c$, the long time limit remains zero.
 
For $\Fex < \Fex^c$, the real part of the correlator is a monotonously decaying function of time, whereas the imaginary part exhibits a maximum.
For $\Fex > \Fex^c$, both the real and the imaginary parts of the tracer
correlator decay faster and faster with increasing $\Fex$ and eventually
start to oscillate.  These oscillations are due to the $\Fex$-dependent 
term $\om$ (see eqs.~(\ref{fex_sjgr}), (\ref{om_fex}),  (\ref{fex_f1})), which dominates at high forces.
The oscillations thus arise for all bath states at larger forces
(see Fig.~\ref{fex_corr_fig}, panels (a-h)).

For the case that the bath is in the arrested state but the coupling between
the tracer and the bath is small (see panels (c-d) of
Fig.~\ref{fex_corr_fig}), the long time limit of the tracer correlator is
equal to zero for all values of $\Fex$. The behaviour of $\phi^s(t)$ is
similar to that for the case of strong probe-bath coupling for $\Fex > \Fex^c$.


Finally, for the case of the liquid bath (panels (g-h) of Fig.~\ref{fex_corr_fig}),
we can see that besides  lowering the value of the intermediate
$\beta$-plateau with increasing $\Fex$ (this effect corresponds to the effect
of decreasing long time limit with increasing $\Fex$ for $\eps > 0$),  also
the time scale of the $\alpha$-process (i.~e.~the final decay from the $\beta$-plateau  to zero) decreases. 
After the  critical value $F^c_{ex}(\eps=0)$ is reached, the $\beta$-plateau  becomes zero and the
 difference between the $\beta$-process and the $\alpha$-process
 disappears. The behaviour for $F_{ex}>F^c_{ex}(\eps=0)$ is similar to that of
 the case $\eps > 0$ (as was already mentioned in this section): the overall time scale of the
 decay decreases and for large enough $\Fex$, the tracer correlator starts to oscillate.

In the next sections, we consider the $\beta$-relaxation  and the $\alpha$-relaxation  regions of the $F_{ex}$-Sj\"ogren model
in more detail and make some quantitative predictions.

\subsection[The $\beta$-correlators]{The $\beta$-correlators}\label{beta_cor_sec}

In this section we consider the behaviour of the correlators around the
$\beta$-relaxation plateau and perform a (non-linear) stability analysis of the arrested/localized part
of the correlator. Here, classical MCT has provided the  deepest insights by
deriving results like the factorization theorem and power-law  relaxation
during the so-called $\beta$-process. The $q$-dependent factorization  theorem
shows that the dynamics on all length scales follows  a single, time-dependent
function, the so-called $\beta$-correlator $G(t)$.  It depends sensitively on the
separation to the MCT bifurcation and  introduces algebraic decay into the
dynamics. We perform the non-linear stability  analysis in order to
investigate the de-localization transition at  finite force in more detail.

\subsubsection{The $\beta$-scaling equation}

We use the ansatz
\ba\label{ans_gst}
\phi^s(t) & = & f^s +G^s(t),\\\label{ans_gt}
\phi(t) & = & f + G(t)
\ea
 with the assumptions that $f^s,\,f$ fulfill the long-time limit equation (\ref{sch_ltlim_eq})  and
 the $\beta$-correlators $G^s(t)$, $G(t)$ are small
\be
|G^s(t)|,\,|G(t)| \ll 1.
\ee
Using the standard steps (partial integration etc.),  we rewrite eq.~(\ref{fex_sjgr}) in the form
\be\label{eq_phis_t_rewr}
\partial_t \phi^s(t) = -\om\,\phi^s(t) + m(t) - \frac{d}{dt}\int_0^t dt'\,
m(t-t')\,\phi^s(t')\, .
\ee 
We insert eqs.~(\ref{ans_gst}), (\ref{ans_gt}) into
(\ref{eq_phis_t_rewr}) and obtain
\begin{widetext}
\be\label{eq_gst_full}
\partial_t G^s(t)  =
-\om\, (f^s + G^s(t)) + v_s\, ({f^s}^* +  {G^s}^*(t)) \, (f+G(t))
    - \frac{d}{dt}\,\int_0^t dt'\, v_s\, ({f^s}^* + {G^s}^*(t-t'))\, (f+G(t-t'))\, (f^s+G^s(t')) 
\ee
\end{widetext}
While eq.~(\ref{eq_gst_full}) has not been solved yet for all relevant cases, a number of
solutions exist and provide insight into the tracer dynamics close to delocalization. 

If we retain  only the terms of the order not higher than linear in $G^s$, $G$ in eq.~(\ref{eq_gst_full}), neglect the time derivative 
and make use of eq.~(\ref{sch_ltlim_eq}),   we obtain 
\ba\nonumber
 G^s(t)\,\left({f^s}^* f - \frac{\om}{v_s}\right)  + {G^s}^*(t)\,
 f\,(1-f^s)  \quad\quad\quad\\\label{eq_beta_cor_1ord}
 +\, G(t)\,({f^s}^* - |f^s|^2) = 0 \quad\quad
\ea

\subsubsection{Factorization theorem for fluid states in the $F_{ex}$-Sj\"ogren model}\label{fex_sj_liq}

Here we want to consider the $F_{ex}$-Sj\"ogren model in the liquid state ($\eps<0$)
for the case that the external force is smaller than its critical value for $\eps=0$. This means that the $\beta$-relaxation plateau 
is non-zero.

We consider eq.~(\ref{eq_beta_cor_1ord}) and choose  $f^s$ to be the long time
limit of the tracer correlator for $\eps=0$. 
So,  $f^s\neq 0$ and  eq.~(\ref{eq_beta_cor_1ord}) can be considered 
as a linear equation for $G^s(t)$ with the given $G(t)$.
Expressing $\om/v_s$ in terms of the known functions of $F_{ex}$, namely
$f^s_1$ and $f^s_2$ and using the long-time limit equation (\ref{sch_ltlim_eq}),  we obtain
\be\label{eq_beta_sc_reg}
-G^s(t)\,{f^s}^*\,f + {G^s}(t)^*\,f^s\,f\,(1-f^s) + G(t)\,|f^s|^2\,(1-f^s) =0
\ee
In terms of $G^s_1(t)$, $G^s_2(t)$, i.~e.~the real and imaginary parts of $G^s(t)$,
eq.~(\ref{eq_beta_sc_reg}) can be easily solved with the result
\ba\label{expr_gs1_g}
G^s_1 (t) & = & h_1\,G(t),\\\label{expr_gs2_g}
G^s_2 (t) & = & h_2\,G(t),
\ea
where
\ba\label{h_1}
h_1 & = & \frac{|f^s|^2\,(2 f^s_1\,{f^s_2}^2 + (2 f^s_1 + {f^s_2}^2 -{f^s_1}^2)\,(1-f^s_1))}{f\,(4 {f^s_2}^2f^s_1\,(1-f^s_1) 
        - (2f^s_1 + {f^s_2}^2 -{f^s_1}^2 )\,({f^s_2}^2 -{f^s_1}^2)  )} \quad \\[2ex]\label{h_2}
h_2 & = & \frac{h_1\,f\,({f^s_2}^2 - {f^s_1}^2) + |f^s|^2\,(1-f^s_1)}{2\,f\,f^s_1\,f^s_2}
\ea
This result generalizes the factorization theorem of MCT to forced probes.
As a check, we  set $F_{ex}=0$ and obtain the well-known result of the Sj\"ogren model \cite{Goetze91,Sjoegren86}:
\be
 G^s = G^s_1 + iG^s_2 = G^s_1 = G\,\frac{1}{v_s f^2}.
\ee

\begin{figure}
\includegraphics[scale=0.65]{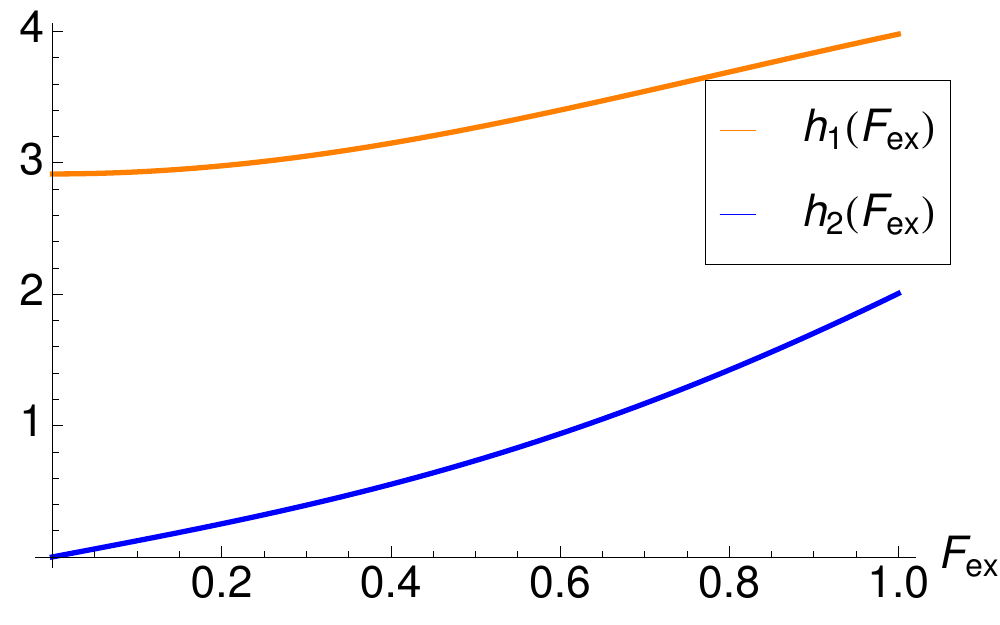}
\caption{\small\label{beta_func_fig} The critical amplitudes 
  (eqs.~(\ref{h_1}), (\ref{h_2})) of the $F_{ex}$-Sj\"ogren model in the fluid state ($v_2=2.0$, $v_s=4.0$, $\Fex^c(\eps=0)=0.61$).}
\end{figure}

The critical amplitudes  $h_1$, $h_2$ are plotted in Fig.~\ref{beta_func_fig} as functions of $F_{ex}$ for fixed
values of $v_2$ and $v_s$. Both functions increase monotonically in the (meaningful) region of the force values $F_{ex}<F^c_{ex}$. 
At $F_{ex}=0$, the function $h_2(F_{ex})$ starts linearly from  zero,
whereas $h_1(\Fex)$ starts quadratically at a non-zero value, as required by symmetry.

\begin{figure}
\begin{tabular}{c}
\includegraphics[scale=0.3]{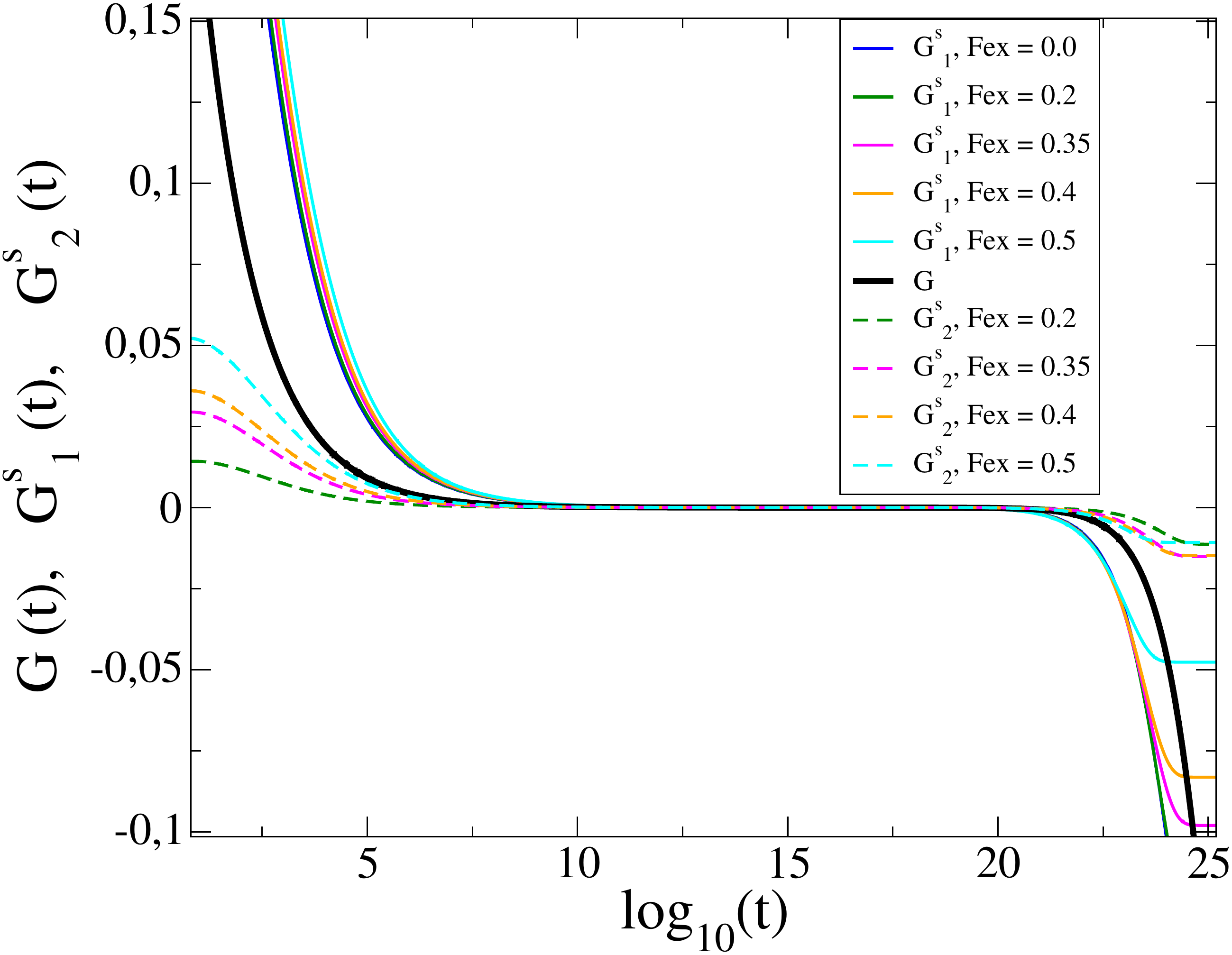}\\[4ex]
\includegraphics[scale=0.3]{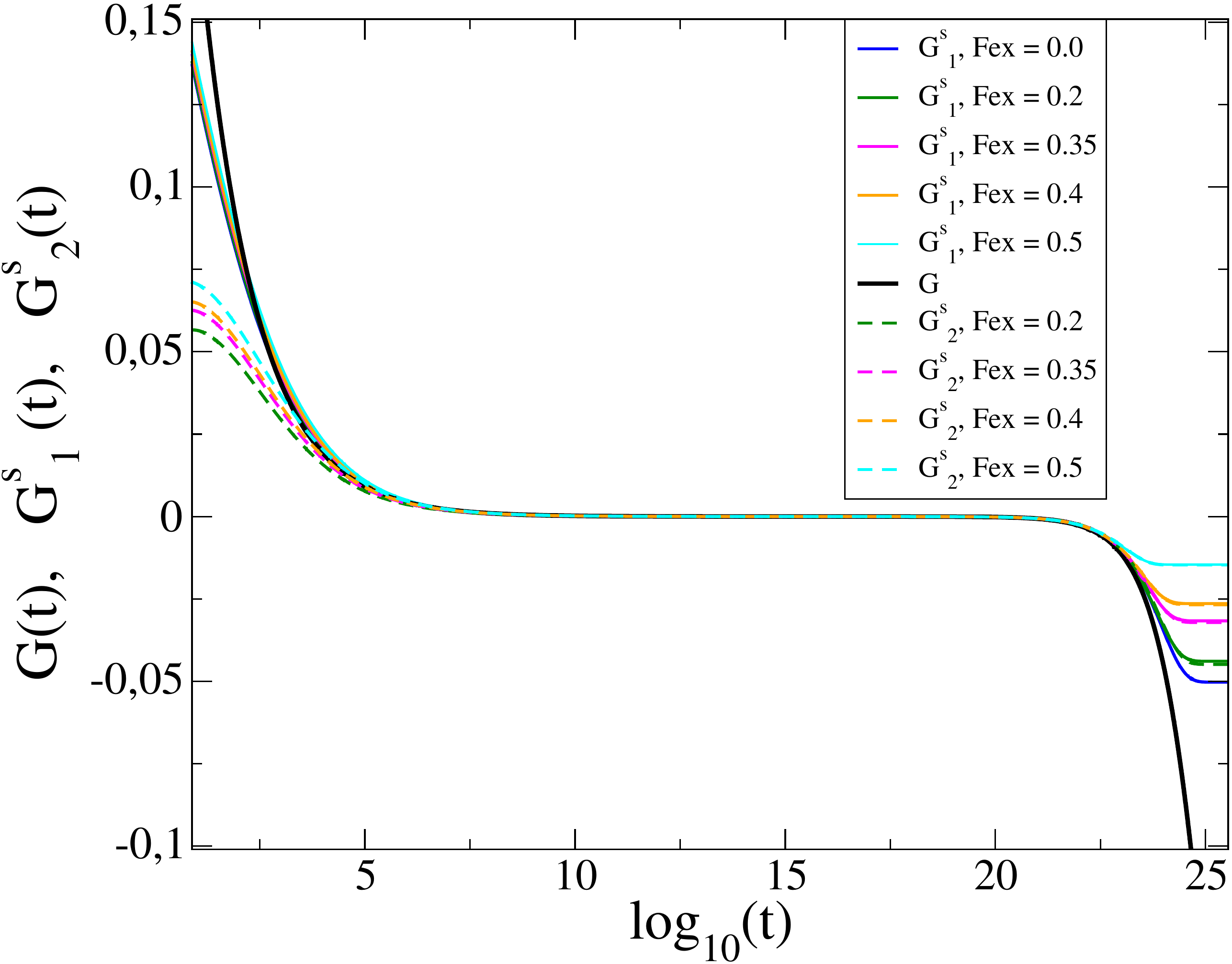}
\end{tabular}
\caption{\small\label{beta_sclg_fig} (Color online). Beta correlators, unscaled (upper panel) and scaled (lower panel) of
 the $F_{ex}$-Sj\"ogren model in the 
fluid state for $\Fex< \Fex^c(\eps=0)$  ($v_2=2.0$, $v_s=4.0$, $\eps=-10^{-11}$, $\Fex^c(\eps=0)=0.61$).}
\end{figure}

To check our results numerically, we plot the $\beta$-correlators (defined in Eqs.~(\ref{ans_gst}, \ref{ans_gt})), first unscaled and then
 scaled according to 
the expressions (\ref{expr_gs1_g}), (\ref{expr_gs2_g}) in Fig.~\ref{beta_sclg_fig}. We see, that  $G^s_1(t)$, $G^s_2(t)$
indeed collapse on the master curve given by $G(t)$, if one is not too far
away from the plateau. This holds for more than ten decades in time in Fig.~\ref{beta_sclg_fig}.

\subsubsection{The critical correlators}\label{crit_cor_sec}

In this section we want to consider the $\beta$-correlators at the critical force value
 $F_{ex}=F^c_{ex}$. So, we set $f^s=0$  in eq.~(\ref{eq_beta_cor_1ord}) 
and obtain
\be\label{eq_crit_cor}
 G^s(t)\,\om  + {G^s}^*(t)\,v_s f =0.
\ee
As can be readily seen, in terms of  $G^s_1(t)$, $G^s_2(t)$ the above equation
can be rewritten as
\ba
J \, \left(
\begin{array}{c}
G^s_1(t)\\
G^s_2(t)
\end{array}\right)  =   0,
\ea
where the matrix $J$ is identical with the one given by eqs.~(\ref{matr_j}), (\ref{matr_i}).
Thus, except from the trivial solution $G^s_1(t)=G^s_2(t)=0$, the  solution of eq.~(\ref{eq_crit_cor}) exists only when the determinant
of the matrix $J$ vanishes, which is exactly the bifurcation condition, as
discussed in Sect.~\ref{bif_an_sec}.
If it is fulfilled, the solution of eq.~(\ref{eq_crit_cor})  is not unique and represents a relationship between $G^s_1(t)$ 
and $G^s_2(t)$. 

 In our case the codimension of the bifurcation, i.~e.~the dimension of the critical space is one, as was shown in 
Sec. \ref{bif_an_sec}. Thus at the bifurcation, 
the correlators $G^s_1(t)$,  $G^s_2(t)$  are  proportional to the critical eigenvector of the stability matrix and thus  to each other.

Considering also the next-to-leading (quadratic) terms in eq.~(\ref{eq_gst_full}) and still assuming $f^s=0$, we obtain
\be\label{eq_gs_g_2ord}
- G^s(t)\,\frac{\om}{v_s} + {G^s}^*(t)\,(f+G(t))  = \frac{d}{dt}\int_0^t dt'\,G^s(t') {G^s}^*(t-t').
\ee

For the $F_{ex}$-$F1$ model, we have to set $f=1$ and $G=0$ in eq.~(\ref{eq_gs_g_2ord}). The left-hand side of eq.(\ref{eq_gs_g_2ord})
vanishes due to the condition (\ref{eq_crit_cor}) and we  obtain the following
equation for the critical $\beta$-correlator:
\be\label{eq_fex_f1_cr}
0  = \frac{d}{dt}\int_0^t dt'\,G^s(t') {G^s}^*(t-t').
\ee
This equation can be solved by means of the power-law ansatz
\ba
G^s(t) = t^x + i\, t^y.
\ea
We get under the integral in eq.~(\ref{eq_fex_f1_cr}) the expression
\be
 t'^x\,(t-t')^x + t'^y\,(t-t')^y + i\,\left[ t'^y\,(t-t')^x - t'^x\,(t- t')^y  \right].
\ee
Using the identity
\be\label{gamma_func_int}
\frac{d}{dt}\int_0^t dt' \,(t-t')^x\,t'^y = t^{x+y}\frac{\Gamma(x+1)\Gamma(y+1)}{\Gamma(x+y+1)},
\ee
where  $\Gamma(x)$ is the gamma function, we see that the imaginary part of the right-hand side in eq.~(\ref{eq_fex_f1_cr}) vanishes.
The choice $x=y=-1/2$  lets also the real part of the  right-hand side in
 eq.~(\ref{eq_fex_f1_cr}) vanish, since then the denominator in eq.~(\ref{gamma_func_int}) diverges.

\begin{figure}
\includegraphics[scale=0.3]{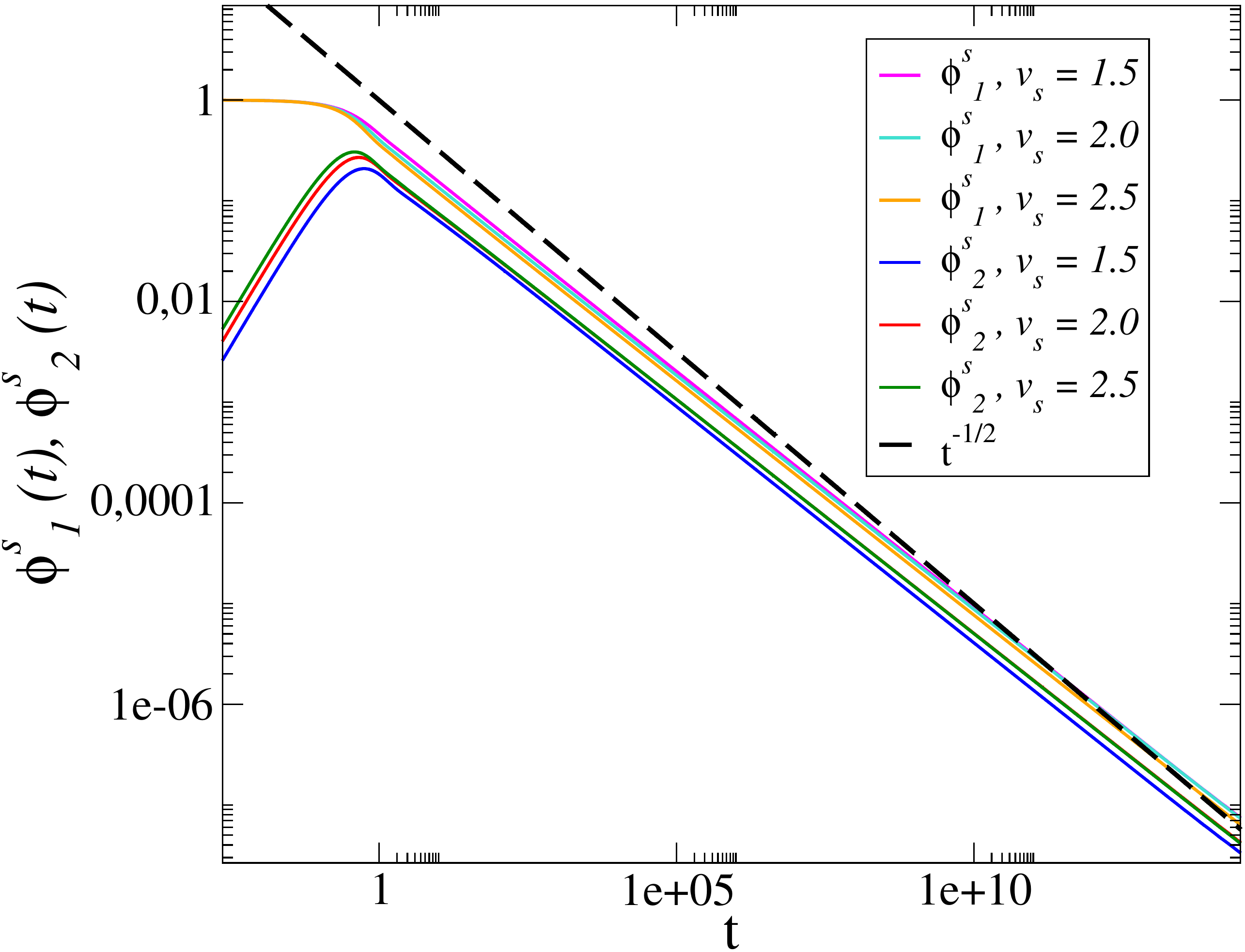}
\caption{\small\label{fex_f1_crit_fig} (Color online). The critical correlators of the $F_{ex}$-$F1$ model for different values of $v_s$.  
The thick black curve represents the asymptotic power law (\ref{fex_f1_cr_sol}).}
\end{figure}

We thus found the {\em power law} solution
\be\label{fex_f1_cr_sol}
G^s(t) = t^{-1/2} + i\,t^{-1/2}
\ee
of the equation (\ref{eq_fex_f1_cr}), which gives the critical $\beta$-correlator of the $\Fex$-F1 model. Fig.~\ref{fex_f1_crit_fig} 
shows the  critical correlators for different  values of the parameter $v_s$. 
We see that the power law (\ref{fex_f1_cr_sol}) indeed holds both  for the real (continuous lines) and the imaginary parts (dashed lines) 
 asymptotically for large times.
The solution (\ref{fex_f1_cr_sol}) can still be multiplied by an arbitrary prefactor. This expresses the
scale invariance of the eq.~(\ref{eq_fex_f1_cr}). The correct prefactor can be found by matching to the initial decay.

If we consider now the  $F_{ex}$-Sj\"ogren model, we get from eqs.~(\ref{eq_gs_g_2ord}), (\ref{eq_crit_cor}) 
\be
 {G^s}^*(t)\,G(t)  = \frac{d}{dt}\int_0^t dt'\,G^s(t') {G^s}^*(t-t').
\ee
This equation has no simple power-law solution, since with the power-law ansatz, the imaginary part of its right-hand side would vanish,
whereas the left-hand side would still have a non-vanishing imaginary part.

\subsection{The $\alpha$-relaxation}\label{fex_sj_liq_2}

We want to consider now the $\alpha$-decay  region of the $\Fex$-Sj\"ogren
model for $\eps<0$  (see panels (g-h) of Fig.~\ref{fex_corr_fig} and the
discussion at the end of Sec.~\ref{ext_force_eff}) in more detail.

The $\alpha$-relaxation  behaviour of the schematic model without the external
force is a well known example of the second  scaling-region of MCT, describing
the final decay of the correlator on time scale $\tau$,  the so-called final,
or  $\alpha$-relaxation time. For the present discussion we  recall, that the
second relaxation step of the correlators asymptotically (for $\eps\rightarrow 0^-$) follows a
scaling-function  \cite{Goetze2009} 
\be\label{eq_alpha_sc_law}
\phi(t) \simeq \tilde{\phi}\left(\frac{t}{\tau(\eps)}\right).
\ee

\begin{figure}
\includegraphics[scale=0.3]{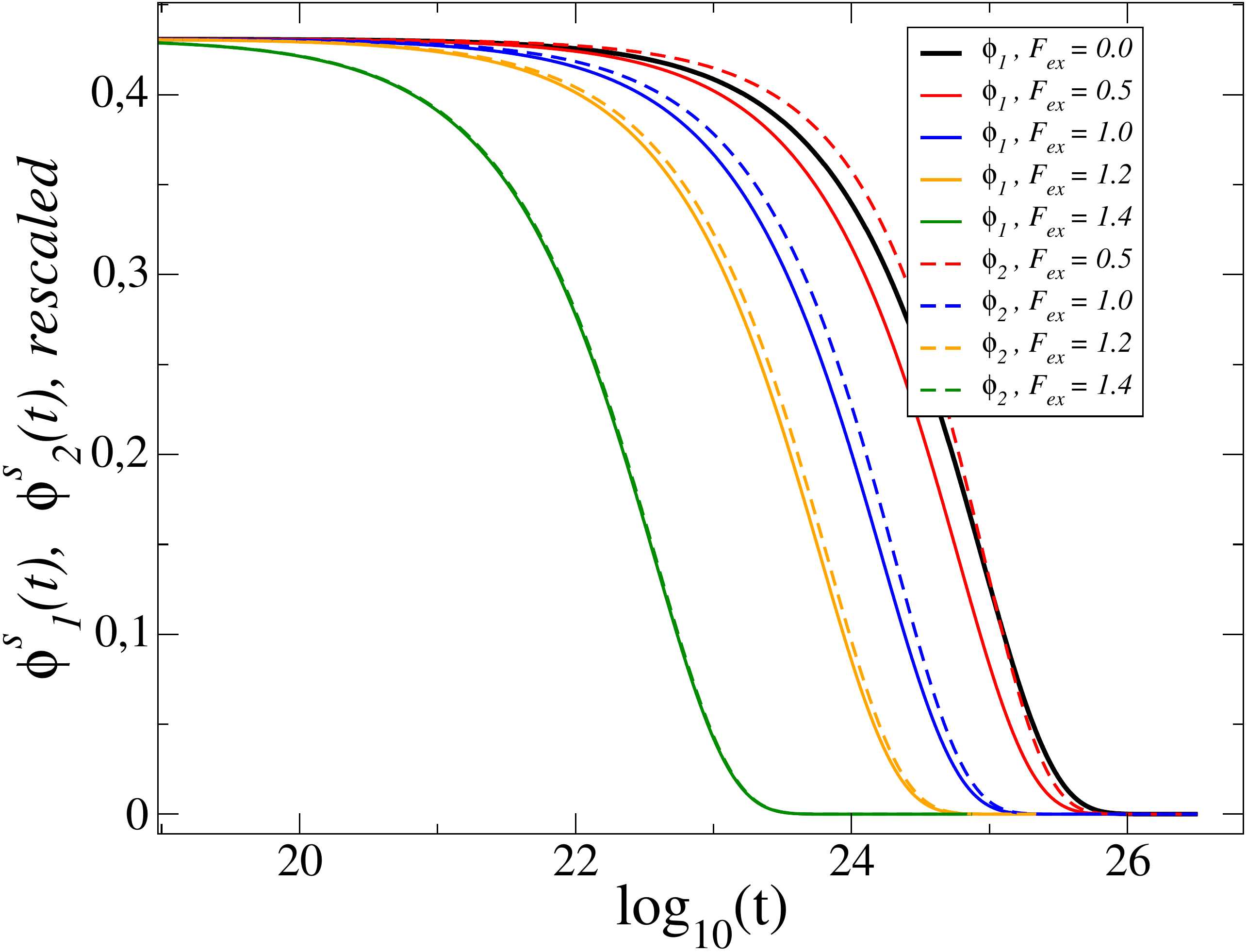}
\caption{\small\label{alpha_sclg_fex_fig} $\alpha$-decay of the correlators, rescaled in the amplitude 
according to their $\beta$-plateau values.
The parameters are $v_s=6.0$ , $v_2=2.0$, $\eps=-10^{-11}$ (this corresponds to $F^c_{ex}=1.445$).}
\end{figure}

As we saw in Sec. \ref{ext_force_eff}, the presence of the external force influences both the time scale of the 
$\alpha$-decay  and the {\em height}  of the $\beta$-plateau. So,
a simple scaling law like (\ref{eq_alpha_sc_law}) cannot work any
more. However, if we rescale the amplitude of the correlator $\phi^s_i$ by the factor 
$\frac{f^s_1(F_{ex}=0)}{f^s_i(F_{ex})}$ ($i=1,\,2$), so that  both
$\phi^s_1$ and $\phi^s_2$  decay from the same ($F_{ex}=0$ real part-) plateau
(see Fig.~\ref{alpha_sclg_fex_fig}), we see that  with increasing $\Fex$, the
shape of both $\phi^s_1$- and $\phi^s_2$-curves  varies slightly. There is also some difference between
  the shape of $\phi^s_1$ and $\phi^s_2$ at the same value of $\Fex$, which decreases 
 with increasing $F_{ex}$ so that for $F_{ex}=1.4$ (slightly below the critical force) $\phi^s_1$ and $\phi^s_2$ almost match.

This observation justifies us to propose the (approximate) generalized ansatz
\be\label{eq_alpha_sc_law_fex}
\phi^s(t) \simeq f^s(\Fex)\, \tilde{\phi^s}\left(\frac{t}{\tau^s(F_{ex}, \eps)}\right),
\ee
suggesting that there is still a universal decay function but accounting for the change in the plateau value. 
 $\tilde{\phi^s}(t)$ is suggested  to be  real, which means that both the real
and the imaginary parts of $\phi^s$ have the same shape. 
The precision of (\ref{eq_alpha_sc_law_fex}) 
can be considered as acceptable  if one realizes that the change of the decay
time scale with increasing $\Fex$  by several orders of magnitude has a much stronger effect than  the minor change in the  shape of the curves.

To determine the tracer $\alpha$-time scale $\tau^s$, we  match
(\ref{eq_alpha_sc_law_fex})  to the $\beta$-decay law (see Sec.~\ref{fex_sj_liq}):
\ba\nonumber
\phi^s_i(t) =   f^s_i + h_i\,G(t) = f^s_i\, \left(1 - \frac{h_i}{f^s_i}\,\left(\frac{t}{\tau(\eps)}\right)^b\right)  = \quad \\
       = f^s_i\,\left(1 - \left(\frac{t}{\tau(\eps)}\,(f^s_i/h_i)^{1/b}\right)^b\right), \quad
\ea
 where the asymptotic form of the bath beta-correlator $G(t)= - \left(t/\tau(\eps)\right)^b$ was used \cite{Goetze91},  to obtain 
\be\label{alpha_sc_fex_eq}
\tau^s_i(\eps,\,F_{ex}) = \tau(\eps)\,\left(\frac{f^s_i}{h_i}(F_{ex}) \right)^{1/b},
\ee
where $i=1,\,2$ corresponds to the real and imaginary part, respectively. This
result will be used in the next section to analyse the low-force behavior of
the tracer friction coefficient in a fluid host.

\subsection{The tracer friction coefficient}\label{fric_coef_sec}

Within the framework of the schematic models, where no wave vector dependence
of the correlators is present, we define  
the friction coefficient increment following our considerations in Sec. \ref{tr_fric_sec}   
as the time integral over the product of the real part of 
the tracer correlator and the bath correlator:
\be\label{del_zeta_schem}
\Delta\zeta_s = \int_0^\infty dt\, \phi^s_1(t)\,\phi(t)
\ee

\subsubsection{$F_{ex}$-F1model}\label{fric_coef_sec_1}

We start by considering first the $F_{ex}$-F1 model, since exact analytical results are available here. 
Eq.~(\ref{fex_f1}) reads in the Laplace space:
\be\label{phis_lapl}
-i\,(z\,\hphis(z)+1) = -(1-i\,F_{ex})\,\hphis(z) -  v_s\,(z\,\hphis(z) + 1)\,\hphis^*(-z).
\ee

We use the following definition of the Laplace transform
\be
\hat f(z) = LT[f(t)] (z) \equiv i\,\int_0^\infty dt\, e^{izt}\,f(t),
\ee
with the properties
\ba
LT\left[\partial_t f(t)\right]\,(z) & = & -i\,(z\,\hat f(z) + f(0)), \quad\quad\quad \\
LT\left[\int_0^t ds\, f(s)\,g(t-s)\right]\,(z) & = & -i\,\hat f(z)\,\hat g(z),\\
LT\left[f^*(t)\right]\,(z) & = & -(\hat f)^* (-z).
\ea

For the calculation of the friction coefficient, only the imaginary part of $\hphis(z=0)$ is of interest, since
$\hphis(z=0)  = - \int_0^\infty dt\,\phi^s_2(t) + i\, \int_0^\infty dt\,\phi^s_1(t)$.
So, we have $\Delta\zeta_s = \Im\{ \hphis(z=0)\}$.  
We set $z=0$ in eq.~(\ref{phis_lapl}), then the product $z\,\hphis$ 
vanishes and  for 
\ba
\hphis_1 &  \equiv & \Re\{\hphis(z=0)\},\\ 
\hphis_2 & \equiv &  \Im\{\hphis(z=0)\}
\ea
we  obtain the following system of equations
\ba
\hphis_1 + F_{ex}\,\hphis_2 & = & -v_s\,\hphis_1\\
1 + v_s\,\hphis_2  & = & \hphis_2 - F_{ex}\,\hphis_1,
\ea
which yields
\be\label{visc_f1}
\Delta\zeta_s = \hphis_2 = \frac{1 + v_s}{F_{ex}^2 + 1 - v_s^2}
\ee
So, we have an exact analytical result 
\footnote{This nice finding, unfortunately,  cannot be transferred to $z\neq 0$, 
except for $F_{ex}=0$ \cite{Goetze91} or for the (physically uninteresting) model 
without complex conjugation in the memory function.} %
 and see that the friction coefficient  exhibits  {\em thinning} behaviour  with increasing $\Fex$. 

As expected from symmetry, $\Del\zeta_s$ starts out quadratically for small external forces for  $v_s<1$, i.~e.~for the case of
 low probe-bath coupling.
 For $v_s>1$, i.~e.~for strong probe-bath coupling, expr.~(\ref{visc_f1}) can be rewritten as
\be\label{del_zeta_f1_glass}
\Del\zeta_s =  \frac{1 + v_s}{F_{ex}^2 - {F^c_{ex}}^2},
\ee
with
\be
F^c_{ex} = \sqrt{v_s^2 - 1}
\ee
(according to eq.~(\ref{cr_Fex_eq}) with $f=1$).
Note that eq.~(\ref{del_zeta_f1_glass})  applies only if $F_{ex} >  F^c_{ex}$, otherwise the tracer is localized 
and $\Del\zeta_s=\infty$ holds.  Due to the identity $F_{ex}^2 - {F^c_{ex}}^2 = (F_{ex} - F^c_{ex})(F_{ex} + F^c_{ex})$,
$\Del\zeta_s$ diverges at $F^c_{ex}$ according to the asymptotic {\em power law}
\be
\Del\zeta_s \sim \frac{1}{F_{ex} - F^c_{ex}}.
\ee

\begin{figure}
\includegraphics[scale=0.31]{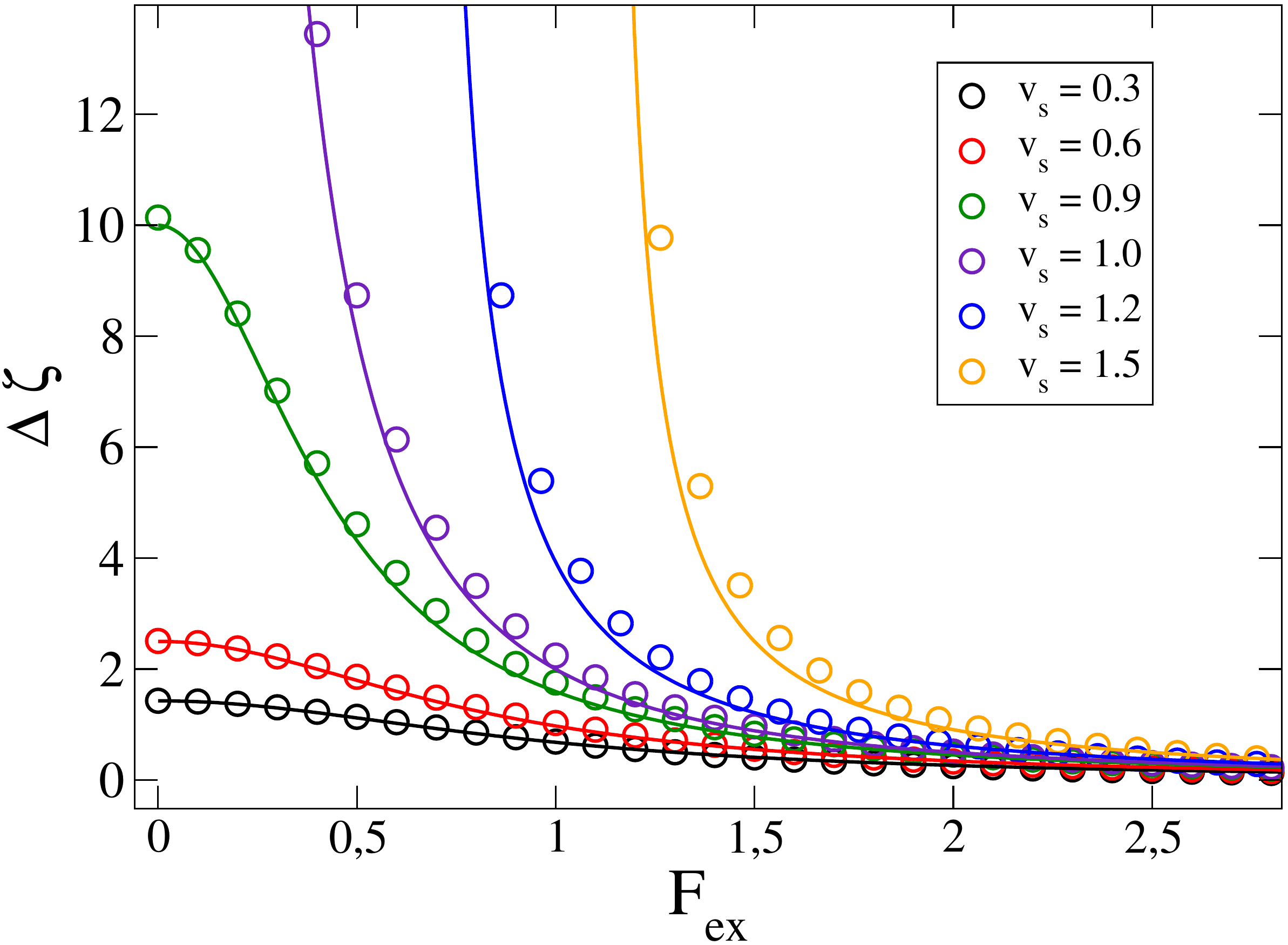}
\caption{\small\label{f1_eta_fex_fig} Friction coefficient increment from the $F_{ex}$-F1 model. The circles show the 
numerical values (from the direct numerical integration of eq.~(\ref{del_zeta_schem})), and  the continuous lines show the analytical 
 values  (calculated from eq.~(\ref{visc_f1})).}
\end{figure}

In Fig.~\ref{f1_eta_fex_fig} we plot the numerical and the analytical values of $\Delta\zeta_s$ for different  $v_s$ 
and observe quite a reasonable agreement. The deviations increase with  $v_s$ (following the general trend in the numerics to 
become unstable at higher values of the probe-bath coupling strength) and  can be considered as a quality
 measure of the numerical procedures used.

\subsubsection{$F_{ex}$-Sj\"ogren model}\label{fric_coef_sec_2}

\begin{figure}
\includegraphics[scale=0.32]{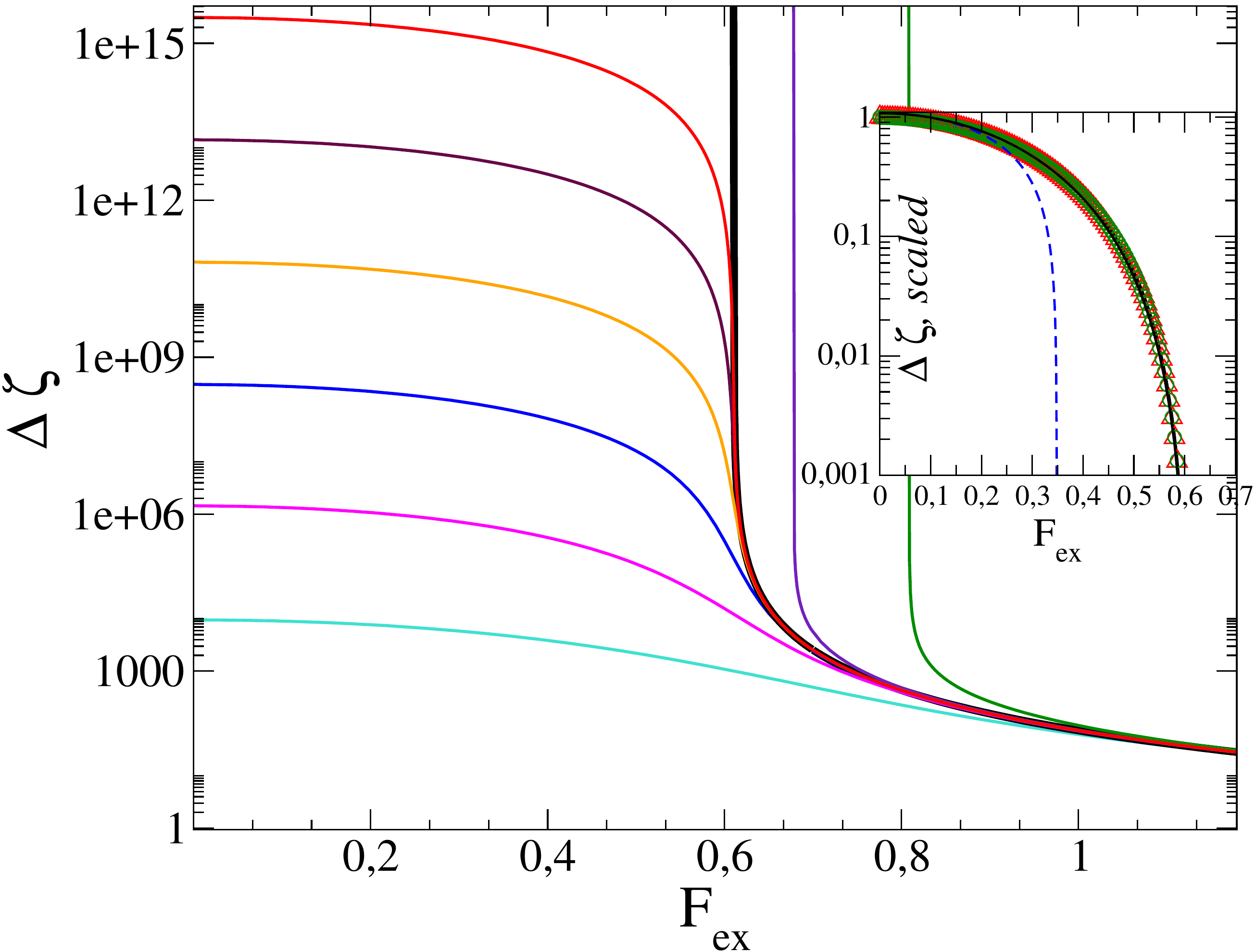}
\caption{\small \label{visc_fex_sjgr_fig} Main panel: friction coefficient increment from
  the $F_{ex}$-Sj\"ogren model ($v_2=2.0$, $v_s=4.0$). Results from the
numerical integration of eq.~(\ref{del_zeta_schem}). The bold
  black line corresponds to $\eps=0$. The values of $\eps$ are $-10^{-3}$,
  $-10^{-4}$,  $-10^{-5}$,  $-10^{-6}$, $-10^{-7}$, $-10^{-8}$, 0, $10^{-4}$, $10^{-3}$ in ascending order. 
  Inset:  data for $\eps = -10^{-9},\, -10^{-8},\,-10^{-7} $ (for $F_{ex} < F^c_{ex}$) scaled
by $\tau(\eps)$  to collapse onto the mastercurve eq.~(\ref{delzeta_scaling}) (continuous line). 
The dashed line corresponds to the small $\Fex$ expansion $\Del\zeta_s(\eps=-10^{-7},\Fex=0)(1-c\,\Fex^2) $ (with $c = 8.35$). }
\end{figure}

For the $F_{ex}$-Sj\"ogren model, the results of the numerical integration of eq.~(\ref{del_zeta_schem}) 
are shown in Fig.~\ref{visc_fex_sjgr_fig} (continuous lines) as function of $\Fex$ for different values of $\eps$.  
We observe {\em thinning} with increasing force and  see that for large forces, all the curves collapse 
on the same  limiting curve. This limiting curve corresponds to the large $F_{ex}$ limit of the schematic model, where the $\om$-term, 
which contains $F_{ex}$ dominates and the  memory term can be neglected so
that one obtains a $1/{F_{ex}^2}$ decay law for $\Del\zeta_s$.
This behavior also agrees with the low-density approximation in the
microscopic MCT equations (see eqs.~(\ref{dzeta_fex}, \ref{f_beta})).

For $\eps\geq0$, the curves diverge at the critical value of force, which increases with $\eps$. On the fluid side for  $\eps < 0$, two
different decay regimes can be distinguished: the strong decay from the initial (linear response) plateau 
for $F_{ex}<F^c_{ex}(\eps=0)$  and the further decay for $F_{ex}>F^c_{ex}(\eps=0)$, which approaches the  
$\eps=0$ limiting curve with increasing force.
The initial decay for $\eps\rightarrow 0^-$ can be analyzed  analytically using the $\alpha$-decay 
law derived in Sec. \ref{fex_sj_liq_2}, since the time integral over the correlators is dominated by the $\alpha$-decay region for
 $F_{ex}<F^c_{ex}(\eps=0)$. 
Relations  (\ref{del_zeta_schem}), (\ref{eq_alpha_sc_law}), (\ref{eq_alpha_sc_law_fex}) lead to
$
\Del\zeta_s \approx f^s_1(\Fex)\, \int_0^\infty dt\, 
 \tilde{\phi^s}\left(\frac{t}{\tau^s_1(\eps,\Fex)}\right)\,\tilde{\phi}\left(\frac{t}{\tau(\eps)}\right).
$
From Fig.~\ref{fex_corr_fig} (g) we see that $\phi^s(t)$ decays much faster than $\phi(t)$ and thus a further approximation is justified, 
where $\phi(t)$ is considered as constant under the integral. This gives the scaling
\be
\label{delzeta_scaling}
\Del\zeta_s(\eps,\Fex) \propto \tau(\eps)\,f^s_1(\Fex)\left(\frac{f^s_1}{h_1}(\Fex)\right)^{1/b},
\ee
where the use of eq.~(\ref{alpha_sc_fex_eq}) was made.
The inset in Fig.~\ref{visc_fex_sjgr_fig} demonstrates the validity of the factorization of the $\eps$- and the $\Fex$-dependence
in $\Del\zeta_s(\eps,\Fex)$: rescaling of the amplitude of the 
$\Del \zeta_s$ vs.~$\Fex$-curves for different $\eps$ leads to their coincidence. To check the $\Fex$-scaling, we plot 
the function $f^s_1(\Fex)\left(\frac{f^s_1}{h_1}(\Fex)\right)^{1/b}$ (circles on Fig.~\ref{visc_fex_sjgr_fig}; 
the value of the bath beta-scaling exponent $b=0.63$ was used) and observe a good agreement with
the results of the direct numerical integration of eq.~(\ref{del_zeta_schem}).  For small $\Fex$, the expansion
 $\Del\zeta_s(\eps,\Fex) = \Del\zeta_s(\eps, \Fex=0)(1-c\,\Fex^2 + \mathcal{O}(\Fex^4))$ is valid with $c = 8.35$ 
(see inset in Fig.~\ref{visc_fex_sjgr_fig}).

\section{Summary and Conclusions}\label{concl_outlook}

The main  objective of this work was to extend the standard  mode-coupling theory for the motion of a tracer particle in a 
dense colloidal suspension near the glass transition to the case, where the  tracer experiences an  external force $F^c_{ex}$, 
which cannot be assumed to be small compared to the  internal interactions of the system. This means, an attempt is made to go beyond the 
linear response regime. We use formally exact generalized Green-Kubo relations and follow the ideas of the integration through 
transients approach \cite{Fuchs2002c}, recently developed for sheared systems.

The presence of the external force leads to a  drastic difference compared  to the linear response case:
 the tracer density correlator becomes complex. This is the consequence of the fact that instead of the unperturbed 
Smoluchowski operator $\Omd_0$, the full operator $\Omd$ containing the external force enters $\phisq$. $\Omd$ turns out to be 
non-hermitian with respect to the equilibrium average, as the consequence of the fact that we consider an {\em open} system.  
Interestingly, in the mode-coupling theory under shear   density correlators depending on advected wavevectors can be defined so as to remain real, as the affine drift motion of the particles can be taken into account rigorously. 
In the present case of force driven microrheology the drift motion results from the particle interactions and manifests itself in 
a phase factor which needs to be calculated, and which turns the correlator complex.

Despite this qualitative difference in the transient structural relaxation, in fluid states an external force gives thinning behavior 
of the friction coefficient akin to shear-thinning in flow. The difference to flow-driven macrorheology becomes evident in the existence 
of a critical force in glass states. At $\Fex^c$  a continuous  {\em bifurcation} transition of the long-time limit of the tracer density 
correlator  occurs. For $F_{ex} > F^c_{ex}$, the long-time limit becomes zero and thus
 the cage surrounding the tracer breaks.   The probe-particle  becomes delocalized and can be pulled through the suspension. 

In Sec. \ref{schem_models} we constructed the arguably most simple schematic models  by considering only two wave vectors
parallel to the external force. Two different models are considered: the ``$F_{ex}$-Sj\"ogren model'', extending the  Sj\"ogren model
  \cite{Sjoegren86} for the tracer coupled to a bath and the ``$F_{ex}$-$F1$ model'', extending the $F1$ model of standard MCT,
 which was used to describe the tracer in a matrix of immobile particles (the Lorentz model).  
The long-time limits and the phase diagrams could be calculated analytically. The bifurcation at $\Fex^c$ was shown to have codimension one.

Numerical and asymptotic solutions of the time-dependent equations of motion of the schematic models enabled predictions of the force
dependence of the tracer friction increment $\Del\zeta_s$.
Generally, the thinning behaviour is observed similar to the shear thinning in macrorheology.
 For the $F_{ex}$-$F1$ model, an exact analytic expression for $\Del\zeta_s$ could be derived, 
showing a power law divergence with the exponent $-1$ at the critical force.
For the $F_{ex}$-Sj\"ogren model, scaling laws could be given for small and large external forces.



\begin{acknowledgments}

We thank M.~Gnann, C.~Harrer, A.~M.~Puertas, and Th.~Voigtmann for discussions.
This work was (partially) funded by the German Science Foundation in SFB 513 and by the German Excellence Initative.
I.~G.~would like to thank Prof.~J.~U.~Sommer and Leibniz Institut f\"ur Polymerforschung Dresden for funding  during 
the completion of this paper.

\end{acknowledgments}

\bibliography{phd_ref,mct}

\end{document}